\documentclass[11pt]{extarticle}
\usepackage[utf8]{inputenc}
\usepackage[fleqn]{amsmath}
\usepackage{gensymb} 
\usepackage{bm}
\usepackage{colortbl}  
\usepackage[table,dvipsnames]{xcolor}    
\usepackage{graphicx, wrapfig, subcaption, setspace, booktabs}
\usepackage{epstopdf}  
\usepackage{subcaption} 
\usepackage{float}
\usepackage{parskip}
\usepackage{empheq}
\usepackage[theorems,skins]{tcolorbox}
\usepackage{xcolor}
\usepackage{tikz}
\usepackage{natbib}
\usepackage{comment}
\usepackage[a4paper, total={14cm, 24cm},]{geometry}
\usepackage{layout}    
\usepackage[colorlinks = true,
            linkcolor = blue,
            urlcolor  = blue,
            citecolor = blue,
            anchorcolor = blue]{hyperref}
\usepackage{tabularray}
\UseTblrLibrary{booktabs}
\usepackage[counterclockwise]{rotating}
\usepackage[graphicx]{realboxes} 
\usepackage{adjustbox} 

\usepackage{lipsum}

\usepackage{empheq}
\usepackage{glossaries}

\newcommand{\Deq}[1]{$D$}
\newcommand{\ddt}[1]{\frac{\text{d}#1}{\text{d} t}}

\newcommand{\wrt}[1]{\text{d}#1}

\definecolor{color1}{HTML}{1B9E77} 
\definecolor{color2}{HTML}{D95F02} 
\definecolor{color3}{HTML}{7570B3} 
\definecolor{color4}{HTML}{E7298A} 
\definecolor{color5}{HTML}{66A61E} 
\definecolor{color6}{HTML}{E6AB02} 
\definecolor{color7}{HTML}{A6761D} 
\definecolor{color8}{HTML}{666666} 
\definecolor{color_green}{HTML}{A9AE19} 
\definecolor{color_racinggreen}{HTML}{1B9569} 
\definecolor{color_TU_blue}{HTML}{006699}
\definecolor{color_TU_lightblue}{HTML}{7FB2CC} 
\definecolor{color_TU_green}{HTML}{379237}

\newacronym{ibm}{IBM}{Immersed Boundary Method}

\begin{document}

\title{\Large\textbf{An immersed boundary method for  
particle-resolved simulations of arbitrary-shaped rigid particles}}
\author{Maximilian Schenk$^{a}$,
Manuel Garc\'ia-Villalba$^{a}$,  Jan Du\v{s}ek$^{b}$, \\ 
Markus Uhlmann$^{c}$ and Manuel Moriche$^{a}$}
\date{\vspace{-2ex}}
\maketitle
\begin{center}
$^{a}$\textit{Institute  of Fluid Mechanics and Heat Transfer, 
	TU Wien, Vienna, Austria} \\
$^{b}$\textit{
Universit\'{e} de Strasbourg, Department of Mechanics,
Institut ICube, Strasbourg, France} \\
$^{c}$\textit{Institute for Water and Environment, Karlsruhe Institute 
	of Technology, Karlsruhe, Germany} \\
\end{center}

\rule{\textwidth}{0.6pt}
\vspace{-0.8cm}

\paragraph{Abstract} \text{ }\\
\vspace{-0.4cm}

The present work extends the direct-forcing immersed boundary method 
introduced by García-Villalba et al. (2023), 
broadening its application from spherical to arbitrarily-shaped particles, 
while maintaining its capacity to address both neutrally-buoyant and light 
objects (down to a density ratio of 0.5). 
The proposed method offers a significant advantage over existing methods 
regarding its simplicity, in particular for the case of neutrally-buoyant 
particles.
Three test cases from the literature are selected for validation: a 
neutrally-buoyant prolate spheroid in a shear flow; a settling oblate spheroid; 
and, finally, a rising oblate spheroid.

\rule{\textwidth}{0.6pt}


\section{Introduction}

Particle-laden flows are commonly found in the environment, such as microplastic 
pollution in the ocean or snow avalanches, as well as in engineered systems like 
chemical reactors and fluidized beds.
These flows are governed by the complex interaction between fluid and particle 
motion, and for small particles in dilute conditions, point-particle models can 
accurately predict the behavior of particles in the flow. 
However, when the size of the particles is comparable to (or larger than) that 
of the smaller scales of the flow, and/or the concentration of particles is not 
small, the predictive capability of the point-particle approach is still low, 
and one needs to resort to particle-resolved (PR) simulations.
Among the several available techniques \citep{uhlmann2023efficient}, 
the \gls{ibm} has gained popularity for PR simulations of particle-laden flows 
because of its computational efficiency, versatility and accuracy
\citep{mittal_immersed_2005,verzicco_immersed_2023}.

The \gls{ibm} was first developed by \cite{peskin_flow_1972} for heart flow 
simulations.
Since then, it has been expanded to tackle a broad spectrum of problems 
\citep{griffith_immersed_2020,mittal_immersed_2021,arranz_fluidstructure_2022}.
Unlike methods with a boundary-conforming mesh, which require complex mesh 
adaptations to account for moving or flexible bodies, \gls{ibm} simplifies 
the process by allowing bodies to be treated with a fixed computational grid.
\citet{sotiropoulos_immersed_2014} and \citet{mittal_origin_2023} present a 
comprehensive review of the available \gls{ibm} approaches,
Among them, a widely used method is the direct-forcing method of 
\citet{uhlmann_immersed_2005} due to its computational efficiency, proven 
accuracy and ease of implementation.
However, in Uhlmann's method, stability considerations lead to a lower 
limit for the density ratio between the solid and fluid phases of approximately 1.2.
Several approaches have been taken to overcome this problem.
\citet{kempe_improved_2012} and \citet{breugem_second-order_2012} independently 
worked on the forcing scheme, each improving numerical stability and enabling 
simulations with a solid/fluid density ratio greater than 0.3.
\citet{schwarz_immersed_2016} further modified the  approach of 
\citet{kempe_improved_2012} by introducing the virtual mass method, adding a 
stabilizing term to both sides of the Newton-Euler equations.
Another modification to the method of \cite{kempe_improved_2012} was proposed 
by \citet{tschisgale_non-iterative_2017, tschisgale_general_2018}.
The most important modification with respect to the original method of Uhlmann
is a different coupling between the fluid and solid equations. 
In addition, the method relies on using a surface layer of mass and, as a 
consequence, for non-spherical bodies, additional terms are required in the 
equations to handle the possible non-overlapping between the centers of mass 
of the particle and the surface layer. 
Recently, \cite{garcia-villalba_efficient_2023} proposed a modification to 
the direct-forcing method described in \cite{uhlmann_immersed_2005} by
incorporating some of the concepts of the solid-fluid coupling from
\cite{tschisgale_non-iterative_2017}.
The method proposed by \cite{garcia-villalba_efficient_2023}
shows no instability for neutrally-buoyant and even moderately-light particles
(with density ratio larger than 0.5), keeping the spirit of Uhlmann's method
regarding its simplicity. 

In this paper, we extend the method described by 
\cite{garcia-villalba_efficient_2023}, which focuses solely on spheres, to 
handle non-spherical particles of arbitrary shape while maintaining the 
advantages of being able to deal with neutrally-buoyant and moderately-light 
objects.
The simplicity of the proposed methodology stands out as a substantial advantage, 
as it eliminates the need for any extra terms related to the possible non-overlap 
between the particle's center of gravity and its surface shell as in 
\cite{tschisgale_non-iterative_2017, tschisgale_general_2018}. 
The simplicity of the formulation is especially evident in the case of 
neutrally-buoyant particles, as will be shown below.
However, the applicability of the methodology is somewhat lower, being limited 
for stability considerations to particles with a density ratio larger than 0.5 
as already discussed by \citet{garcia-villalba_efficient_2023}. 

The paper is organized as follows. 
The governing equations and the coupling between the fluid and solid phases are 
provided in \S~\ref{sec:methodology} together with the full description of the 
flow solver and the spatial discretization.
In \S~\ref{sec:validation} we present a set of validation cases:
i) a neutrally-buoyant prolate spheroid in shear flow,
ii) a settling oblate spheroid in an unbounded domain and 
iii) a light ascending oblate spheroid in an unbounded domain.
The paper concludes with final remarks in \S~\ref{sec:conclusion}. 


\section{Methodology} \label{sec:methodology}

We consider the interaction between particles of arbitrary shape and uniform
density distribution, $\rho_p$, with a Newtonian fluid of constant kinematic
viscosity, $\nu$, and constant density, $\rho_f$.
Further, we assume the flow to be incompressible and the particles to be rigid
bodies.
We extend the method proposed by \cite{garcia-villalba_efficient_2023} to handle
particles of arbitrary shape by i) tracking the rotation of the particles
employing a quaternion-based formulation and ii) solving the angular momentum
equation of the particle in a body-fixed reference frame, similar to
\cite{moriche_single_2021}.

\subsection{Governing equations}
The Navier-Stokes equations for an incompressible flow read 
\begin{align}
    \label{eq:NSE:cont}
    \nabla \cdot \bm{u}&=0, \\
    \label{eq:NSE:mom}
    \frac{\partial \bm{u}}{\partial t}+ \left( \bm{u} \cdot \nabla \right) \bm{u}
     &= - \nabla p + \nu \nabla^2 \bm{u} + \bm{f} ,
\end{align}
where $\bm{u}$ is the fluid velocity, $p$ is the kinematic pressure (i.e.
pressure divided by fluid density) and $\bm{f}$ is a volume force term which
represents the presence of a body (in our case a particle).
The volume force $\bm{f}$ is formulated as a coupling force to impose the
no-slip, no-penetration boundary condition on the surface of the particle.
Consistent with the original method proposed by \cite{uhlmann_immersed_2005},
the equations are applied over the entire domain, $\Omega$, including the fluid,
$\Omega_f$, and the space occupied by the particle, $S$.

Particle motion is driven by hydrodynamic and buoyancy forces, thus it is
governed by the Newton-Euler equations with the corresponding terms
\begin{align}
    \label{eq:LM}
    \rho_p V_p \ddt{\bm{u}_p}
      &=\rho_f \oint_{\partial S} \bm{\tau} \cdot \bm{n}\, \wrt{\sigma}
       + \left( \rho_p - \rho_f \right) V_p \mathbf{g} \,, \\
    \label{eq:AM}
    \ddt{\mathbf{H}_{p}} 
    &= \rho_f \oint_{\partial S} \bm{r} \times \left(\bm{\tau }\cdot \bm{n}\right) 
    \wrt{\sigma} \,,
\end{align}
where $V_p$ is the volume of the particle, $\bm{u_p}=(u_{px},u_{py},u_{pz})$ is
the velocity of its center of mass,
$\bm{\tau}=-p\bm{I}+ \nu \left( \nabla \bm{u}+ \nabla \bm{u}^T \right)$ is the
hydrodynamic stress tensor ($\bm{I}$ is the identity tensor), $\bm{n}$ is the
unit normal vector pointing towards the fluid, and $\bm{g}$ is the gravitational
acceleration.
Further, we define the angular momentum of the particle with respect to the
center of mass $\bm{H}_p=\bm{I}_p\bm{\omega}_p^T$ from the inertia tensor of the
particle, $\bm{I}_p$, and the angular velocity, 
$\bm{\omega}_p=(\omega_{px},\omega_{py},\omega_{pz})$, and the position vector
$\bm {r}$ = $\bm{x}$ - $\bm{x_p}$ of any point in the body, $\bm{x}$, relative
to the center of mass, $\bm{x}_p$.
The surface integrals in (\ref{eq:LM}) and (\ref{eq:AM}) can be related to the
forcing term through Cauchy's principle
\citep[see][Appendix B]{uhlmann_immersed_2005}, as shown below
\begin{align}
   \label{eq:stress_int}
   \oint_{\partial S} \bm{\tau} \cdot \bm{n}\wrt{\sigma}
     &= - \int_S \bm{f}\, \wrt{\bm{x}}+\ddt{ }\int_S \bm{u} \,\wrt{\bm{x}} \,,\\
   \label{eq:stress_ang_int}
   \oint_{\partial S} \bm{r} \times \left( \bm{\tau} \cdot \bm{n} \right) 
	\wrt{\sigma} &=  -\int_S \left( \bm{r} \times \bm{f} \right) 
	\wrt{\bm{x}} 
      + \ddt{ }\int_S \left( \bm{r} \times \bm{u} \right) \, \wrt{\bm{x}} \,. 
\end{align}
The second term of the right-hand side of (\ref{eq:stress_int}) can be 
expressed as
\begin{equation}
    \ddt{ }\int_S \bm{u} \, \wrt{\bm{x}} = V_p \ddt{\bm{u}_p} \, , 
\end{equation}
and further assuming that the flow inside of the particle follows its 
rigid-body motion, we can express the second term in the right-hand 
side of (\ref{eq:stress_ang_int}) as
\begin{equation}
    \label{eq:rbm} 
    \ddt{ }\int_S ( \bm{r} \times \bm{u} ) \, \wrt{\bm{x}}
     =\frac{1}{\rho_p}\ddt{\mathbf{H}_{p}}\,.
\end{equation}
Considering all the assumptions detailed above, equations (\ref{eq:LM}) and
(\ref{eq:AM}) can be rewritten as
\begin{align}
    \label{eq:LMmid} 
    \left(1 - \frac{\rho_f}{\rho_p}\right) V_p \ddt{\bm{u}_p}
       &= - \frac{\rho_f}{\rho_p} \int_{S} \bm{f} \wrt{\bm{x}}
        + \left(1 - \frac{\rho_f}{\rho_p}\right) V_p \mathbf{g} \,, \\
    \label{eq:AMmid}
    \left(1 - \frac{\rho_f}{\rho_p}\right)\ddt{\mathbf{H}_{p}}
       &= - \rho_f \int_S ( \bm{r} \times \bm{f} ) \, \wrt{\bm{x}}   \,.
\end{align}
%
%
\subsection{Coupling condition}
\label{sec:fluid_solid_coupling}
Following \cite{garcia-villalba_efficient_2023}, we combine the convective, 
pressure and viscous terms from equation (\ref{eq:NSE:mom}) into the variable 
$\bm{rhs}$
\begin{equation}
    \bm{f}=\frac{\partial \bm{u}}{\partial t} - \bm{rhs} \label{eq:f} \; ,
\end{equation}
and integrate equation (\ref{eq:f}) in time as originally proposed by 
\cite{tschisgale_non-iterative_2017}
\begin{equation}
    \int_{t_{n-1}}^{t_n} \bm{f} \wrt{t} = \int_{t_{n-1}}^{t_n} \left( 
	\frac{\partial \bm{u}}{\partial t} - \bm{rhs} \right)\wrt{t} = 
    \bm{u}^n - \bm{u}^{n-1} - \int_{t_{n-1}}^{t_n} \bm{rhs} \, \wrt{t} 
	\label{eq:intf} \,,
\end{equation}
where $\bm{u}^n$ is the fluid velocity at the time instant $t_n$.
Equation (\ref{eq:intf}) can be rewritten as
\begin{equation}
    \int_{t_{n-1}}^{t_n} \bm{f} \wrt{t} = \bm{u}^n- \bm{\tilde u} \quad  
	\forall \bm{x} \in S \,,
\end{equation}
where $\bm{\tilde u}$ is an estimated velocity obtained explicitly without 
considering the presence of the body, hence ignoring the forcing term $\bm{f}$
\begin{equation}
    \bm{\tilde u} = \bm{ u}^{n-1} + \int_{t_{n-1}}^{t_n} \bm{rhs} 
	\,\wrt{t} \label{eq:prim_u}\, .
\end{equation}
Under the rigid-body assumption, the particle velocity and the fluid velocity 
at any interior point at the end of the interval ($t^n$) are related through
\begin{equation}
    \bm{u}^n=\bm{u}_p^n + \bm{\omega}_p^n \times \bm{r}^n \quad  \forall 
	\bm{x} \in S \label{eq:rigid_u} \, .
\end{equation}

Integrating the solid-phase equations, (\ref{eq:LMmid}) and (\ref{eq:AMmid}), 
over the time interval $[t_{n-1},t_{n}]$ results in
\begin{align}
    \left(1 - \frac{\rho_f}{\rho_p}\right) V_p \left( \bm{u}_p^{n} - 
	\bm{u}_p^{n-1} \right) &= - \frac{\rho_f}{\rho_p} \int_{S} 
	\int_{t_{n-1}}^{t_n} \bm{f} \wrt{t} \wrt{\bm{x}} + \left( 1- 
	\Delta t \frac{\rho_f}{\rho_p} \right) V_p \mathbf{g} \, , 
	\label{eq:int_time_u} \\
    \left( 1- \frac{\rho_f}{\rho_p} \right) \left( \bm{H}_{p}^{n} 
	- \bm{H}_{p}^{n-1} \right) &= - \rho_f \int_S \int_{t_{n-1}}^{t_n} 
	( \bm{r} \times \bm{f} ) \, \wrt{t} \, \wrt{\bm{x}} \, 
	\label{eq:int_time_omega} \,,
\end{align}
where $\Delta t = t_{n}-t_{n-1}$.
An additional benefit of the present methodology is linked to the exact 
temporal integration of eq.(\ref{eq:AMmid}) leading to eq.
(\ref{eq:int_time_omega}). 
As we do not need to compute the time derivative of $\bm{H}_p$, then 
we do not need to deal with  the cross-term 
$\bm{\omega}_p \times \bm{H}_{p}$
that appears when taking time derivates in non-inertial reference frames.
The treatment of equation (\ref{eq:int_time_omega}) is simplified by using 
a body-fixed reference frame, so that the inertia tensor is constant.
In addition, the axes of the body-fixed reference frame are aligned with the
principal axes of inertia of the body, leading to a diagonal inertia tensor. 
These two steps are not a requirement of the method and could be performed 
differently if needed for some particular reason.
The right-hand side of eq. (\ref{eq:int_time_omega}) is available in the 
laboratory frame, so that
it needs to be expressed in the body-fixed frame by using the rotation 
matrix, $\mathcal{R}$, of the previous step. This is because the new 
orientation of the body is not yet known \citep{moriche_single_2021}. 
More specifically, we define the rotation matrix as a function of the 
particle orientation in terms of the quaternion $\mathbf{q}$ as follows:
\begin{equation}
    \mathcal{R}^n=\mathcal{R}(\mathbf{q}^n),
\end{equation}
where the entries in the matrix $\mathcal{R}$ are given in 
\eqref{eq:rotation_matrix}.
Hence, the angular momentum equation in a body-fixed reference frame can 
be written as:
\begin{equation}
\left( 1- \frac{\rho_f}{\rho_p} \right) \bm{I}_{p,b} \left( \bm{\omega}_{p,b}^{n} 
	- \bm{\omega}_{p,b}^{n-1} \right) = - \rho_f \mathcal{R}^{n-1}
	\left[ \int_S \int_{t_{n-1}}^{t_n} ( \bm{r} \times \bm{f} ) \,
	\wrt{t} \, \wrt{\bm{x}} \right] \,,\label{eq:int_time_omega_body}
\end{equation}
where the subscript $b$ stands for the body-fixed reference frame.

Next, we follow \cite{tschisgale_non-iterative_2017} to express the double 
integrals on the right-hand side of equations (\ref{eq:int_time_u}) and 
(\ref{eq:int_time_omega_body}) in terms of the particle kinematics at the 
end of the interval and of the preliminary velocity
\begin{align}
    \int_{S} \int_{t_{n-1}}^{t_n} \bm{f} \wrt{t} \, \wrt{\bm{x}} &= 
	\int_S ( \bm{u}_p^n + \bm{\omega}_p^n \times \bm{r}- \bm{\tilde u}) 
	\wrt{\bm{x}}= V_p \bm{u}_p^n-\int_S \bm{\tilde u} \,\wrt{\bm{x}} 
	\label{eq:int_order_u} \,, \\
    \int_S \int_{t_{n-1}}^{t_n} ( \bm{r} \times \bm{f} ) \wrt{t} \, 
	\wrt{\bm{x}} &= \int_S \left[ \bm{r} \times ( \bm{u}_p^n + 
	\bm{\omega}_p^n \times \bm{r}- \bm{\tilde u})\right] \wrt{ \bm{x}}= 
	\frac{\bm{I}_p}{\rho_p}\bm{\omega}_p^n-\int_S \bm{r} \times 
	\bm{\tilde u} \, \wrt{ \bm{x}} \label{eq:int_order_omega} \,.
\end{align}
Combining the last steps by substituting eqs. 
(\ref{eq:int_order_u})-(\ref{eq:int_order_omega}) into eqs. 
(\ref{eq:int_time_u}) and (\ref{eq:int_time_omega_body}) and rearranging, 
we obtain
\begin{align}
\bm{u}_p^n&=\left( 1- \frac{\rho_f}{\rho_p} \right)\bm{u}_p^{n-1}+\frac{1}{V_p} 
	\frac{\rho_f}{\rho_p} \int_S \bm{\tilde u} \, \wrt{\bm{x} }+ \Delta t 
	\left( 1- \frac{\rho_f}{\rho_p} \right) \bm{g} \label{eq:int_final_u} \,,\\
\bm{\omega}_{p,b}^n&=\left( 1- \frac{\rho_f}{\rho_p} \right)\bm{\omega}_{p,b}^{n-1}
	+\bm{I}_{p,b}^{-1} \rho_f \mathcal{R}^{n-1} \left[ \int_S \bm{r} \times 
	\bm{\tilde u} \, \wrt{ \bm{x} } \right] \label{eq:int_final_omega} \,.
\end{align}
The studies by \cite{yu_direct-forcing_2007} and \cite{moriche_single_2021} show 
that forcing throughout the volume enclosed by the particle, $S$, is superior 
to forcing only on its surface, $\partial S$, since it leads to a velocity field 
that fulfills the rigid-body assumption more closely.
Furthermore, if the distribution of Lagrangian markers throughout the volume 
is sufficiently uniform and homogeneous, the integrals in \eqref{eq:int_final_u} 
and \eqref{eq:int_final_omega} can be approximated to second-order accuracy by 
using discrete sums, as numerically verified by 
\cite{garcia-villalba_efficient_2023}.
We use this fact as an opportunity to apply forcing inside the particle
at practically no additional cost in the interpolation step of the \gls{ibm}
(please refer to \ref{sec:app:ibm_surface_volume} for an analysis
of the computational costs of surface- and volume-forcing approaches, and
the implications that these IBM-related costs would have when considering
many particles).
Considering the above, the two integrals in \eqref{eq:int_final_u} 
and \eqref{eq:int_final_omega} can be readily approximated using discrete sums
\begin{align}
    \int_S \bm{\tilde u} \, \wrt{\bm{x}} &\approx \sum_{l=1}^{N_l} 
	\bm{\tilde U}_l \Delta {V}_l \; ,  \label{eq_disc_sum_u}\\
    \int_S \bm{r} \times \bm{\tilde u} \, \wrt{\bm{x}} &\approx 
	\sum_{l=1}^{N_l} \bm{R}(\bm{X}_l) \times \bm{\tilde U}_l 
	\Delta {V}_l \; \label{eq_disc_sum_omega} ,
\end{align}
where $\Delta {V}_l$ and $\bm{R}(\bm{X}_l)=\bm{X}_l-\bm{x_p}$ are the discrete 
volume and the position relative to the particle center, respectively, of the
$l$th Lagrangian marker, and $\bm{\tilde U}_l$ denotes the preliminary 
velocity of the $l$th marker, that will be used later to compute the 
forcing term.
Substituting the discrete sums into equations (\ref{eq:int_final_u}) and 
(\ref{eq:int_final_omega}), we finally obtain 
\begin{align}
\bm{u}_p^n&=\left( 1- \frac{\rho_f}{\rho_p} \right)\bm{u}_p^{n-1}+\frac{1}{V_p} 
	\frac{\rho_f}{\rho_p} \sum_{l=1}^{N_l} \bm{\tilde U}_l \Delta {V}_l + 
	\Delta t \left( 1- \frac{\rho_f}{\rho_p} \right) \bm{g} 
	\label{eq:final_u} \,, \\
\bm{\omega}_{p,b}^n&=\left( 1- \frac{\rho_f}{\rho_p} \right)\bm{\omega}_{p,b}^{n-1}
	+\bm{I}_{p,b}^{-1} \rho_f \mathcal{R}^{n-1}\left[ \sum_{l=1}^{N_l} 
	\left( \bm{R}(\bm{X}_l) \times \bm{\tilde U}_l \right) \Delta {V}_l  \right] 
	\label{eq:final_omega}\,.
\end{align}
The velocity updates shown in equations \eqref{eq:final_u} and 
\eqref{eq:final_omega} exhibit the same stability properties as in 
\cite{garcia-villalba_efficient_2023}, i.e. they are unstable 
for $\rho_p/\rho_f \leq 0.5$.

Finally, for the case of neutrally-buoyant particles ($\rho_p/\rho_f=1$), 
equations (\ref{eq:final_u})-(\ref{eq:final_omega}) are reduced to
\begin{align}
    \bm{u}_p^n&=\frac{1}{V_p} \sum_{l=1}^{N_l} \bm{\tilde U}_l \Delta {V}_l 
	\, , \label{eq:final_nb_u} \\
    \bm{\omega}_{p,b}^n&=\bm{I}_{p,b}^{-1} \rho_f \mathcal{R}^{n-1}
	\left[ \sum_{l=1}^{N_l} \left( \bm{R}(\bm{X}_l) \times \bm{\tilde U}_l 
	\right) \Delta {V}_l  \right] \, . \label{eq:final_nb_omega}
\end{align}
\subsection{Flow solver and spatial discretization}
We solve the governing equations in a similar way as in the original method 
proposed by \cite{uhlmann_immersed_2005}, except for the use of the extended 
fluid-solid coupling technique, which is described in detail 
in \S~\ref{sec:fluid_solid_coupling}.
To ensure continuity, we solve equations \eqref{eq:NSE:cont} and 
\eqref{eq:NSE:mom} by means of a projection method (\cite{brown_accurate_2001}).
Spatial discretization is done with second-order finite differences on a staggered, 
uniform grid and time marching is performed with a low-storage semi-implicit 
three-stage Runge-Kutta scheme, where linear terms are treated implicitly, and 
non-linear terms explicitly.
Denoted as $\bm{x}^\beta_{ijk}$, the grid points' positions refer to the staggered 
grid associated with the velocity component $u_\beta$, where $\beta$ takes 
values of 1, 2 and 3.

Lagrangian quantities are represented by uppercase letters.
To uniformly distribute $N_L$ points throughout the volume of a non-spherical 
particle, we generate a 3D Vorono\"{i} tessellation and use Lloyd's algorithm 
\citep{lloyd1982least} to iterate their coordinates until we achieve
convergence.
These positions are represented as $\bm{X}_l \: \forall \: S$, 
with $ 1 \leq l \leq N_L $.
A discrete volume $\Delta V_l$ is assigned to each point so that the particle's 
total volume is the sum of these discrete volumes.
The procedure for obtaining the distribution of points is illustrated in 
\cite{moriche_single_2021}.

Quantities are transferred between Lagrangian and Eulerian grids using the 
regularized delta function, $\delta_h$, as described in 
\cite{peskin_immersed_2002} and defined in \cite{roma_adaptive_1999}.
For the $k$-th Runge–Kutta stage, we first compute the preliminary velocity 
$\bm{\tilde u}$ by advancing the momentum equation ignoring the presence of 
the particle and without considering the continuity constraint 
\begin{equation}
    \bm{\tilde u} = \bm{u^{k-1}}+\Delta t \left( 2 \alpha_k \nu \nabla^2 
	\bm{u}^{k-1} - 2 \alpha_k \nabla p^{k-1} - \gamma_k \left( (\bm{u} \cdot 
	\nabla) \bm{u} \right)^{k-1} - \xi_k  \left( (\bm{u} \cdot \nabla) 
	\bm{u} \right)^{k-2} \right) \,,
\end{equation}
where the coefficients $\alpha_k$, $\gamma_k$, $\xi_k$ ($1\leq k \leq 3$) are 
those used by \cite{rai_direct_1991}.
Afterwards, the preliminary velocity is transferred from the Eulerian to the 
Lagrangian grid
\begin{equation}
    \tilde U_{\beta,l}=\sum_{ijk} \tilde{u}_\beta(\bm{x}_{ijk}^\beta)\delta_h 
	\left( \bm{x}_{ijk}^\beta - \bm{X}_l^{k-1} \right) \Delta x^3,  
	\quad \forall l; 1 \leq \beta \leq 3 \,.
\end{equation}
In the original method by \cite{uhlmann_immersed_2005}, the subsequent step 
is computing the force volume term using the particle velocity of the 
previous stage $k-1$.
Instead, in the current method, the next step is to compute the present 
particle velocity, $\bm{u}_p^k$, and the present angular velocity of the 
particle, $\bm{\omega}_{p,b}^k$, computed in the body-fixed reference frame
\begin{align}
    \bm{u}_p^k&=\left( 1- \frac{\rho_f}{\rho_p} \right)\bm{u}_p^{k-1}+
	\frac{1}{V_p} \frac{\rho_f}{\rho_p} \sum_{l=1}^{N_l} \bm{\tilde U}_l 
	\Delta {V}_l + 2 \alpha_k \Delta t \left( 1- \frac{\rho_f}{\rho_p} 
	\right) \bm{g} \, ,\label{eq:numerical_upk} \\
    \bm{\omega}_{p,b}^k&= \left( 1- \frac{\rho_f}{\rho_p} \right) 
	\bm{\omega}_{p,b}^{k-1}+\bm{I}_{p,b}^{-1} \rho_f \mathcal{R}^{k-1}
	\left( \sum_{l=1}^{N_l} \bm{R}(\bm{X}_l^{k-1}) \times \bm{\tilde U}_l 
	\Delta {V}_l  \right) \, .
\end{align}
The appearance of $2\alpha_k$ in equation (\ref{eq:numerical_upk}) is due to 
the derivation of equation (\ref{eq:final_u}) using a generic time step 
$\Delta t$, whereas the time step associated with the k-$th$ stage 
is $2\alpha_k \Delta t$.
For a neutrally-buoyant case, we can replace these equations with
\begin{align}
    \bm{u}_p^k&=\frac{1}{V_p} \frac{\rho_f}{\rho_p} \sum_{l=1}^{N_l} 
	\bm{\tilde U}_l \Delta {V}_l \, ,\label{eq:numerical_upk_nb} \\
    \bm{\omega}_{p,b}^k&=\bm{I}_{p,b}^{-1} \rho_f \mathcal{R}^{k-1}
	\left( \sum_{l=1}^{N_l} \bm{R}(\bm{X}_l^{k-1}) 
	\times \bm{\tilde U}_l \Delta {V}_l  \right) \, .
\end{align}
Using these velocities of the rigid body, we can now determine the 
new desired velocity
\begin{equation}
    \bm{U}^{(d)}_l=\bm{u}_p^k+ \left({\mathcal{R}^{k-1}}\right)^T 
	\left( \bm{\omega}_{p,b}^k \times \bm{R}(\bm{X}_{l,b})\right) 
	\quad \forall l \, . \label{eq:u_desired}
\end{equation}
Please note that in (\ref{eq:u_desired}) the rotation matrix is 
computed with the quaternion from the previous Runge-Kutta stage.
Following this, the subsequent steps of the original fluid phase 
method from \cite{uhlmann_immersed_2005} remain unchanged. 
As a result, the following operations can be performed in sequence:
\begin{gather}
    \bm{F}_l=\frac{\bm{U}^{(d)}_l-\bm{\tilde U}_l}{\Delta t} 
	\quad \forall l \,, \\
    f_\beta^k(\bm{x}_{ijk}^\beta)=\sum_{l=1}^{N_l} F_{\beta,l} 
	\delta_h(\bm{x}_{ijk}^\beta-\bm{X}_l^{k-1})\Delta V_l 
	\quad \forall \, i, j ,k; 1 \leq \beta \leq 3 \,, \\
    \nabla^2 \bm{u}^*-\frac{\bm{u}^*}{ \alpha_k \Delta t \nu}
	=-\frac{1}{ \alpha_k \nu} \left(
	\frac{\bm{\tilde u}}{\Delta t} + \bm{f}^k \right) + 
	\nabla^2\bm{u}^{k-1} \,, \\
    \nabla^2\phi=\frac{\nabla \cdot \bm{u}^*}{2\alpha_k\Delta t}
	\,, \\
    \bm{u}^k=\bm{u}^*-2 \alpha_k \Delta t \nabla \phi \,, \\
    p^k = p^{k-1}+\phi-\alpha_k \Delta t \nu \nabla^2\phi \,,
\end{gather}
where $\phi$ is the pseudo-pressure
Then, we determine the new position of the particle's center 
of mass, $\bm{x}_p^k$, update the quaternions, $\bm{q}^k$, and also 
update the rotation matrix, $\mathcal{R}^k= \mathcal{R}(\bm{q}^k)$, 
which is defined in \ref{sec:app:rot_mat} along with 
the matrix $\mathbf{Q}^k=\mathbf{Q}(\bm{\omega}^k_{p,b})$.
%
%
\begin{gather}
    \bm{x}_p^k=\bm{x}_p^{k-1}+\alpha_k \Delta t \left( \bm{u}_p^{k} 
	+ \bm{u}_p^{k-1} \right) \,, \\
    \frac{\bm{ \tilde q}^k-\bm{q}^{k-1}}{\Delta t} = \gamma_k
	\frac{1}{2}\bm{Q}(\bm{\omega}_{p,b}^{k-1})\bm{q}^{k-1}+\xi_k 
    \frac{1}{2} \bm{Q}(\bm{\omega}_{p,b}^{k-2})\bm{q}^{k-2} \,, \\
    \bm{q}^k= \frac{\bm{\tilde q}^k}{|\bm{ \tilde q}^k|} \,, \\
    \mathcal{R}^k= \mathcal{R}(\bm{q}^k) \,, \\
    \mathbf{Q}^k=\mathbf{Q}(\bm{\omega}^k_{p,b}) \, .
\end{gather}
And as a final step, the position of the Lagrangian markers is updated: 
\begin{equation}
    \bm{X}^k_l=\bm{x}_p^k+({\mathcal{R}^{k}})^T \bm{R}(\bm{X}_{l,b}) \,.
\end{equation}
Thus, we conclude the Runge–Kutta stage with this calculation.


\section{Validation} \label{sec:validation}

\begin{figure}[b!]
    \centering
    \begin{subfigure}[b]{0.3\textwidth}
        \centering
        \includegraphics[width=\textwidth]{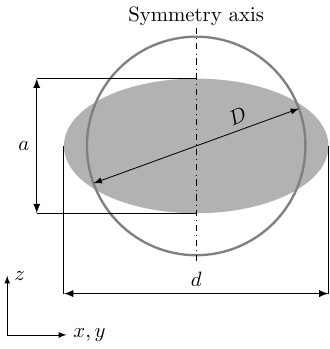}
        \includegraphics[width=\textwidth]{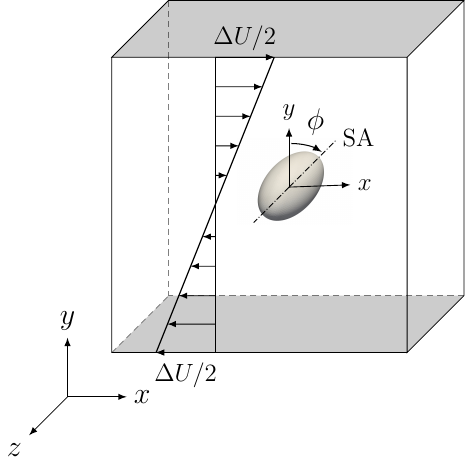}
        \label{fig:ellipsoid_deq}
    \end{subfigure}
    \hfill
    \begin{subfigure}[b]{0.3\textwidth}
        \centering
        \includegraphics[width=\textwidth]{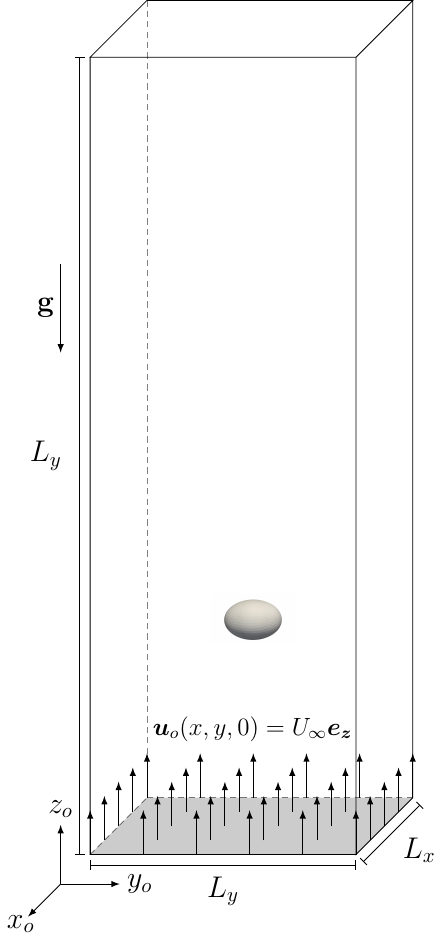}
        \label{fig:settling_domain}
    \end{subfigure}
    \hfill
    \begin{subfigure}[b]{0.3\textwidth}
        \centering
        \includegraphics[width=\textwidth]{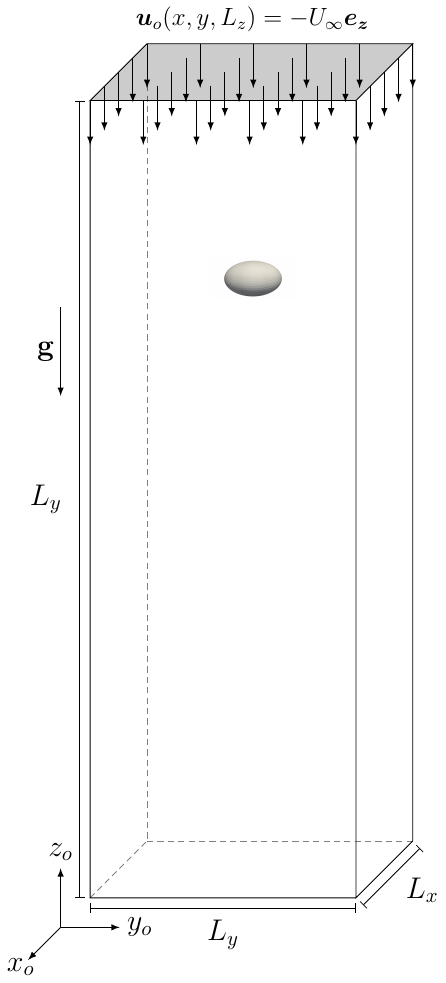}
        \label{fig:rising_domain}
    \end{subfigure}
    \begin{picture}(0,0)
        \put(-462.5,295){(a)}
        \put(-462.5,145){(b)}
        \put(-300,295){(c)}
        \put(-140,295){(d)}
    \end{picture}
    \captionsetup{width=0.9\textwidth}
    \caption{a) Sketch of a spheroid with $\chi=d/a=2$ and equivalent 
	sphere $D=d/\chi^{1/3}$. b) Computational setup for neutrally-buoyant 
	prolate spheroid in shear flow with a visual representation of the angle, 
	$\phi$. Sketch of the computational set-up for c) settling and 
	d) rising particles}
    \label{fig:validation_setup}
\end{figure}
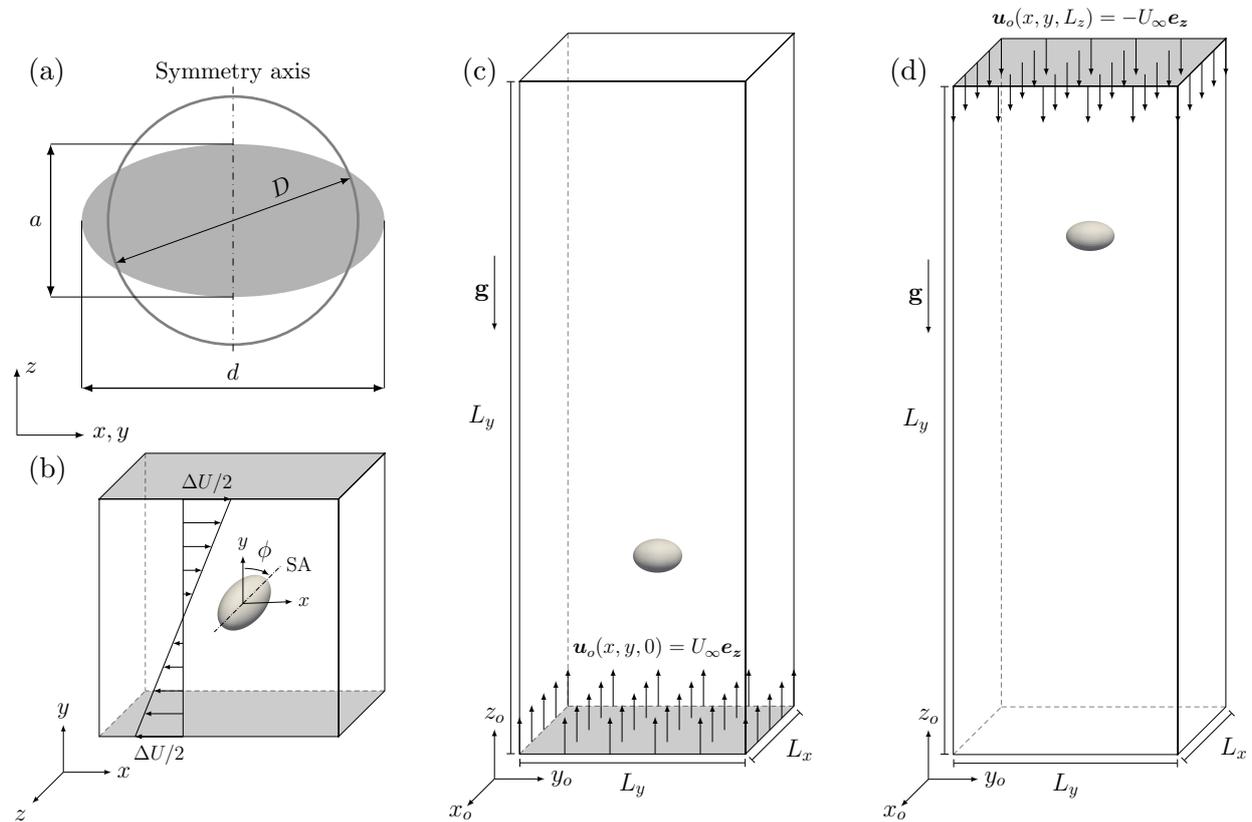
In this section, we present the validation of the method proposed above.
We have selected distinct scenarios, but in all of them we use particles of
spheroidal shape with an aspect ratio $\chi=d/a$ ($\chi < 1$ for prolate
spheroids; $\chi > 1$ for oblate spheroids), where $d$ represents the equatorial
diameter and $a$ is the symmetry axis length.
Additionally, $D$ denotes the equivalent diameter, describing the diameter of a
sphere with the same volume (see Figure \ref{fig:validation_setup}a). 
In the absence of an exact solution, we will use for comparison results obtained 
with a highly-accurate spectral/spectral-element solver
\citep{chrust_numerical_2013}, as has been done in previous works
\citep{uhlmann_motion_2014,rettinger_comparative_2017,moriche_single_2021}.

In section (\ref{sec:settling}) and (\ref{sec:ascending}), the relative errors of
converged values are computed for any quantity $\phi$ as
\begin{equation}
    \varepsilon(\phi) = \frac{| \phi - \phi_{ref}|}{\phi^*_{ref}},
    \label{eq:rel_error}
\end{equation}
where the subscript ``ref'' denotes a value taken from the reference data.
Since some quantities to be compared are of very small amplitude (e.g. the 
horizontal velocity in Table \ref{tab:settling_stobl} below),we use an 
order unity quantity to form $\phi^*_{ref}$ for the purpose of normalization 
(e.g. the settling particle velocity, cf. details below and discussion 
in \cite{uhlmann_motion_2014}, p. 234).

\subsection{Neutrally-buoyant prolate spheroid in shear flow}
\label{sec:jeffery_orbit}
As a first validation case, we study the rotational behavior of a
neutrally-buoyant prolate spheroid in a shear flow.
Following \citet{tschisgale_general_2018}, we select a cubic domain of side
length $L=6.4D$ with periodic boundary conditions along the streamwise and 
spanwise directions ($x,\,z$). 
Dirichlet boundary conditions are imposed at the top and bottom boundaries, 
namely $(u,v,w)=\pm (\Delta U/2,0,0)$ at $y=\pm L/2$.
The prolate spheroid has an aspect ratio $\chi=1/2$.
It is initially placed with the center of mass at the center of the
computational domain, with its symmetry axis aligned with the $y$-axis.
The density ratio is set to 1, and the kinematic viscosity, $\nu$, and the shear
rate $\dot{\gamma}=\Delta U/L$ are adjusted so that 
$Re_{\dot \gamma}=\dot \gamma D^2/\nu=5/32$.
Note that the analytical solution is available for Stokes flow ($Re=0$), while
the present calculations are performed at low but finite $Re$.
The calculations are performed using $128^3$ grid points, corresponding to a
grid resolution of $D/\Delta x=20$.
We initialize the flow with the velocity field ($u(y)=\dot \gamma y,v=w=0$) that
would be the solution in the absence of the particle.  

Under Stokes flow conditions, i.e. $Re\ll1$, \cite{jeffery_motion_1922} derived
an analytical solution of the problem, known as a Jeffery orbit.
This solution is also discussed in detail in \cite{guazzelli_physical_2012}.
When the symmetry axis of the spheroid is contained in an $xy$-plane, the
rotational motion of the spheroid is periodic and uniquely determined by the
angle $\phi(t)$ and its time derivative $\dot{\phi}(t)$.
Figure \ref{fig:validation_setup}b shows a sketch of the problem and a visual
representation of the angle $\phi$. 
Following the notation of \cite{guazzelli_physical_2012} the period is given by
\begin{equation}
    T\dot{\gamma} =  2\pi \left( \chi +1/\chi \right), 
\end{equation}
and the angle of rotation and angular velocity as a function of time are
\begin{align}
  \phi &= \arctan \left( \frac{1}{\chi} \tan \left( 
	\frac{\dot \gamma t}{\chi +\frac{1}{\chi}} \right) \right) \,,\\
  \dot \phi&=\frac{\dot \gamma}{1+\chi^2}\left( \chi^2 \sin^2 \phi 
	+ \cos^2 \phi \right) \,.
\end{align}

\begin{table}[b!]
    \centering    %
    \captionsetup{width=0.9\textwidth}
    \caption{Parameters and results for the calculations of the 
	neutrally-buoyant prolate spheroid in a shear flow. Errors are 
	computed with respect to the analytical solution by \cite{guazzelli_physical_2012}.}
    \scalebox{0.8}{
    \begin{booktabs}{colspec={|rcc|cc|cc|cc|},row{odd}={gray!20},row{1}={bg=gray!40,fg=black}}
    \midrule
    Method  & $D/\Delta x$ & CFL & $T\dot{\gamma}$ & $\varepsilon$ & $\dot \phi_{max}$ & $\varepsilon$ & $\dot \phi_{min}$ & $\varepsilon$ \\
    \midrule
    Analytical &  -    &   -   & 15.707 & -      & 0.8    & -      & 0.2    &       \\
    \midrule 
    IBM        & 20.0   & 0.1  & 16.118 & 0.0260 & 0.8114 & 0.0114 & 0.1848 & 0.0152 \\
    IBM        & 20.0  & 0.01  & 15.946 & 0.0152 & 0.8035 & 0.0035 & 0.1925 & 0.0075 \\
    IBM        & 20.0 & 0.001  & 15.842 & 0.0085 & 0.7983 & 0.0016 & 0.1964 & 0.0036 \\
    \midrule
    \end{booktabs}}
    \label{tab:jeffery_orbit}
\end{table}

\begin{figure}[t!]
    \centering
    \includegraphics[width=0.9\textwidth]{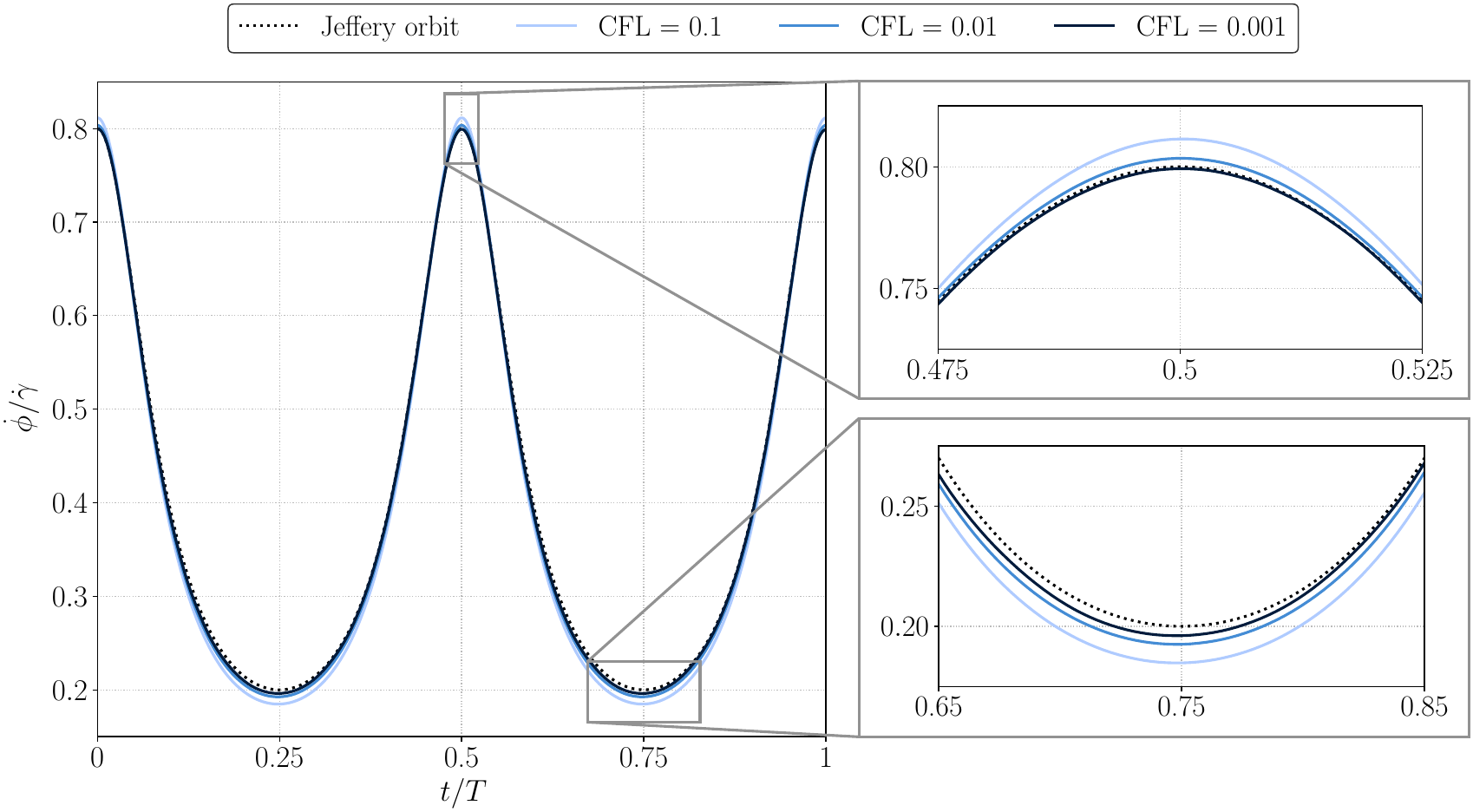}
    \captionsetup{width=0.9\textwidth}
    \caption{Time history of $\dot \phi$ obtained with the proposed method for 
	different CFL-numbers with a resolution of $D/\Delta x=20$ and the 
	analytical solution for a Jeffery orbit 
	(\cite{guazzelli_physical_2012})}
    \label{fig:jeffery_results}
\end{figure}

We have performed three simulations varying the time step so that the CFL
changes from $0.1$ to $0.001$ as reported in Table \ref{tab:jeffery_orbit}.
Figure \ref{fig:jeffery_results} shows the time evolution of the angular
velocity, $\dot{\phi}$.
The agreement is very good for the three calculations and the error decreases
with decreasing time step, as evident by the insets of the figure in the regions
of the local maximum and the local minimum.
A quantitative comparison is reported in Table \ref{tab:jeffery_orbit},
including the period, $T\dot{\gamma}$, and its relative error, along with the
maximum and minimum values of the angular velocity, $\dot{\phi}$, and their
absolute errors.
As the CFL-number decreases, the errors gradually decrease converging to the
analytical values, with the possible exception of the local maximum that seems
to converge to a value that is slightly below the analytical one.
This minor discrepancy may be attributed to the slight differences between the
current configuration and the reference solution, namely finite $Re$ and bounded
domain.

\subsection{Settling spheroid in unbounded domain}
\label{sec:settling}

Let us now analyze the settling of a single oblate spheroid in an unbounded
domain (see Figure \ref{fig:validation_setup}c).
Dimensional analysis shows that the problem is governed by three non-dimensional
parameters, namely the aspect ratio, $\chi$, the density ratio between the
particle and the fluid, $\tilde \rho=\rho_p/\rho_f$, and the Galileo number
$Ga={U_g D}/{\nu}$, where $U_g=\sqrt{|\tilde \rho-1|gD}$ is the gravitationally
scaled velocity.
We evaluate two different Galileo numbers for a spheroid with $\chi=1.5$ and
$\tilde \rho=2.15$, resulting in two different regimes, namely, the
steady-oblique regime and the vertical-periodic regime.
As reference we follow the work of \cite{moriche_single_2021} (referred in the
following as MUD21) to compare the present results to results from a 
spectral/spectral-element method (henceforth called SEM) and from another IBM
that employs the original fluid-solid coupling of \cite{uhlmann_immersed_2005}.
As in MUD21, the cuboidal domain has side lengths $L_x = L_y =6.1D$, a height of
$L_z =18.3D$ and we select grid resolutions with $D/\Delta x=21$ and
$D/\Delta x=42$. 
This results in a grid size of $(128 \times 128 \times 384)$ points for the
lower resolution and $(256 \times 256 \times 768)$ points for the higher
resolution.
We impose at the bottom boundary a Dirichlet boundary condition with velocity
$\bm{u}(x,y,0)=U_{\infty}\bm{e}_z$.
At the top boundary, we impose an advective boundary condition and periodic
boundary conditions are imposed in the lateral directions ($x$ and $y$).
Following the approach used in previous studies (\cite{uhlmann_motion_2014}),
we adjust the Reynolds number based on $U_\infty$ so that it is as close as
possible to the mean settling Reynolds number based on the terminal velocity of
the particle, therefore obtaining long integration intervals without incurring
high computational costs.

\subsubsection{Steady-oblique regime}
\label{sec:settling_stobl}

First, we consider the steady-oblique regime. 
This regime is ideal for benchmarking because of the simplicity of its resulting
kinematics and the narrow range of $Ga$ in which it appears, making it a hard
test to reproduce, but to some extent easy to analyze. 
For the oblate spheroid used here ($\chi=1.5, \tilde \rho = 2.14$) this regime is
observed for $Ga \approx [138,160]$ (see MUD21, Figure 6).
Therefore, we choose $Ga=152$ and the time-step is adjusted in order to keep the
$CFL \approx0.3$.
In this regime, settling particles follow a steady, inclined trajectory,
maintaining a constant tilt angle relative to the vertical axis.
The angle of the trajectory,
\begin{equation}
    \alpha = \arctan \left( \frac{u_{pH}}{u_{pV}} \right),
\end{equation}
is defined by the non-dimensional vertical velocity, $u_{pV}=u_{pz}/U_g$, and
the non-dimensional horizontal velocity, $u_{pH}=\bm{u}_p\cdot\bm{e}_{pH}/U_g$,
where the unit vector $\bm{e}_{pH}=(u_{px},u_{py},0)/\sqrt{u_{px}^2+u_{py}^2}$
indicates the horizontal direction.
Interestingly, the tilting angle, $\phi$, of the spheroid is not exactly the
same as the angle of the trajectory. 
Therefore, we use this latter angle as an additional measure for validation.
Table \ref{tab:settling_stobl} includes the vertical, $u_{pV}$, and the
horizontal, $u_{pH}$, velocities, the trajectory angle, $\alpha$, and the
tilting angle, $\phi$, obtained with the IBM proposed in this work and the
differences/errors with respect to the reference data.
Note that MUD21 reports results for IBM simulations where forcing is applied in
the entire volume of the particle and IBM simulations where forcing is applied
in the surface of the particle. 
Both are included in the table.
The data shows that the proposed method is able to capture this regime
adequately.
Comparing the results of the current IBM method with the SEM from MUD21, we
observe a steady convergence under grid refinement, indicating the reliability
of the method.
Additionally, we observe that both velocity components and both angles are 
slightly over-predicted compared to the SEM reference data.
Similar trends are observed for both volume-forcing IBM (-FI) cases from MUD21.
Overall, the errors remain within an acceptable range compared to the errors of
the IBM results from MUD21. 

\begin{table}[h]
    \centering    %
    \captionsetup{width=0.9\textwidth}
    \caption{Parameters and results for the steady-oblique regime for a settling 
	oblate spheroid. Reference data taken from \cite{moriche_single_2021}.
	The corresponding SEM values are used for normalization ($\phi^*_{ref}$ 
	in eq. \eqref{eq:rel_error}), expect for $\varepsilon_H$, where we 
	use $\phi^*_{ref}=u_{p,V}^\mathrm{{SEM}}$.}
    \begin{adjustbox}{max width=\textwidth}  
    \begin{booktabs}{colspec={|rlcc|cc|cc|cc|cc|},row{odd}={gray!20},row{1}={bg=gray!40,fg=black}}
    \midrule
    Method  & Work  & $\frac{D}{\Delta x}$ & Forcing &  $u_{pV}$ & $\varepsilon_V$ & $u_{pH}$ & $\varepsilon_H$ & $\alpha(\degree)$ & $\varepsilon_\alpha$ & $\phi(\degree)$ & $\varepsilon_\phi$ \\
    \midrule
    SEM     & MUD21 (B15-M075-SEM)   & -       & -       & -1.063 & -      & 0.0714 & -      & 3.842 & -      & 5.318 & -       \\
    \midrule
    IBM     & MUD21 (B15-M075-24)    & 21   & Surface & -1.048 & 0.0141 & 0.0616 & 0.0092 & 3.364 & 0.1244 & 4.676 & 0.1207 \\
    IBM     & MUD21 (B15-M075-24-FI) & 21   & Volume  & -1.050 & 0.0122 & 0.0735 & 0.0002 & 4.004 & 0.0421 & 5.710 & 0.0737 \\
    IBM     & MUD21 (B15-M075-48)    & 42   & Surface & -1.054 & 0.0084 & 0.0657 & 0.0054 & 3.566 & 0.0718 & 4.890 & 0.0805 \\
    IBM     & MUD21 (B15-M075-48-FI) & 42   & Volume  & -1.065 & 0.0019 & 0.0757 & 0.0004 & 4.066 & 0.0583 & 5.600 & 0.0530 \\
    \midrule 
    IBM     & Current                & 21   & Volume  & -1.092 & 0.0276 & 0.0793 & 0.0074 & 4.164 & 0.0838 & 5.687 & 0.0693 \\
    IBM     & Current                & 42   & Volume  & -1.086 & 0.0216 & 0.0781 & 0.0063 & 4.111 & 0.0700 & 5.674 & 0.0669 \\
    \midrule
    \end{booktabs}
    \end{adjustbox}
    \label{tab:settling_stobl}
\end{table}
\subsubsection{Vertical-periodic regime}
\label{sec:settling_vertper}
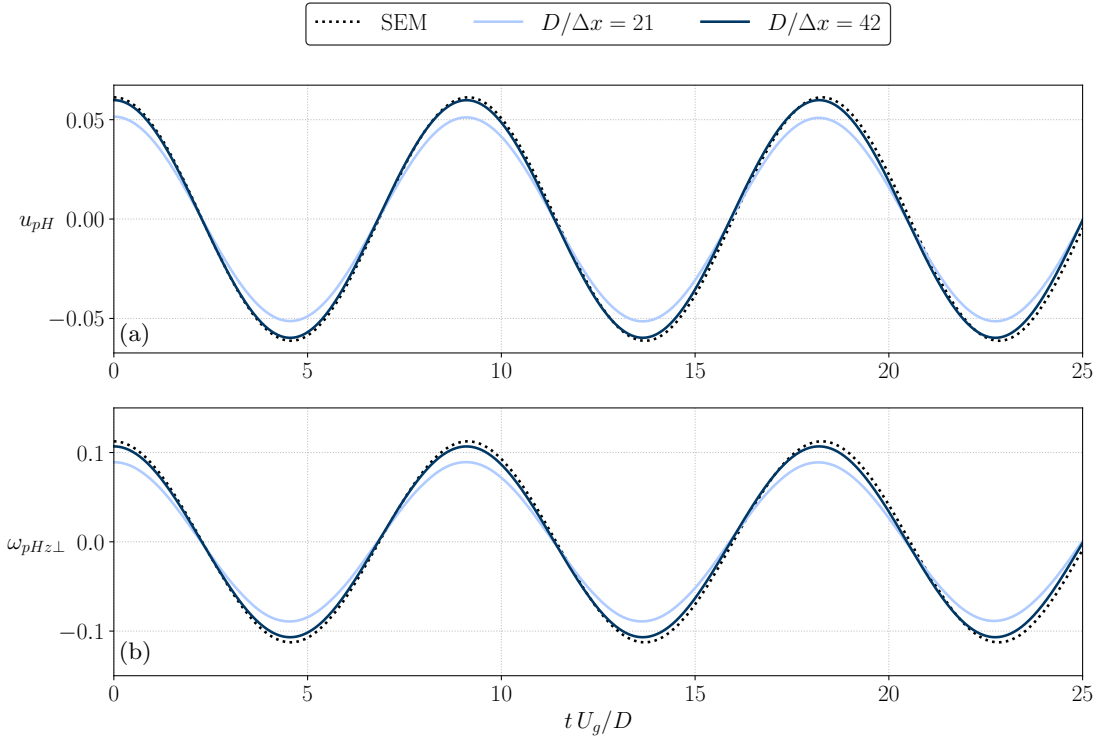
\begin{figure}[b!]
    \centering
    \includegraphics[width=0.9\textwidth]{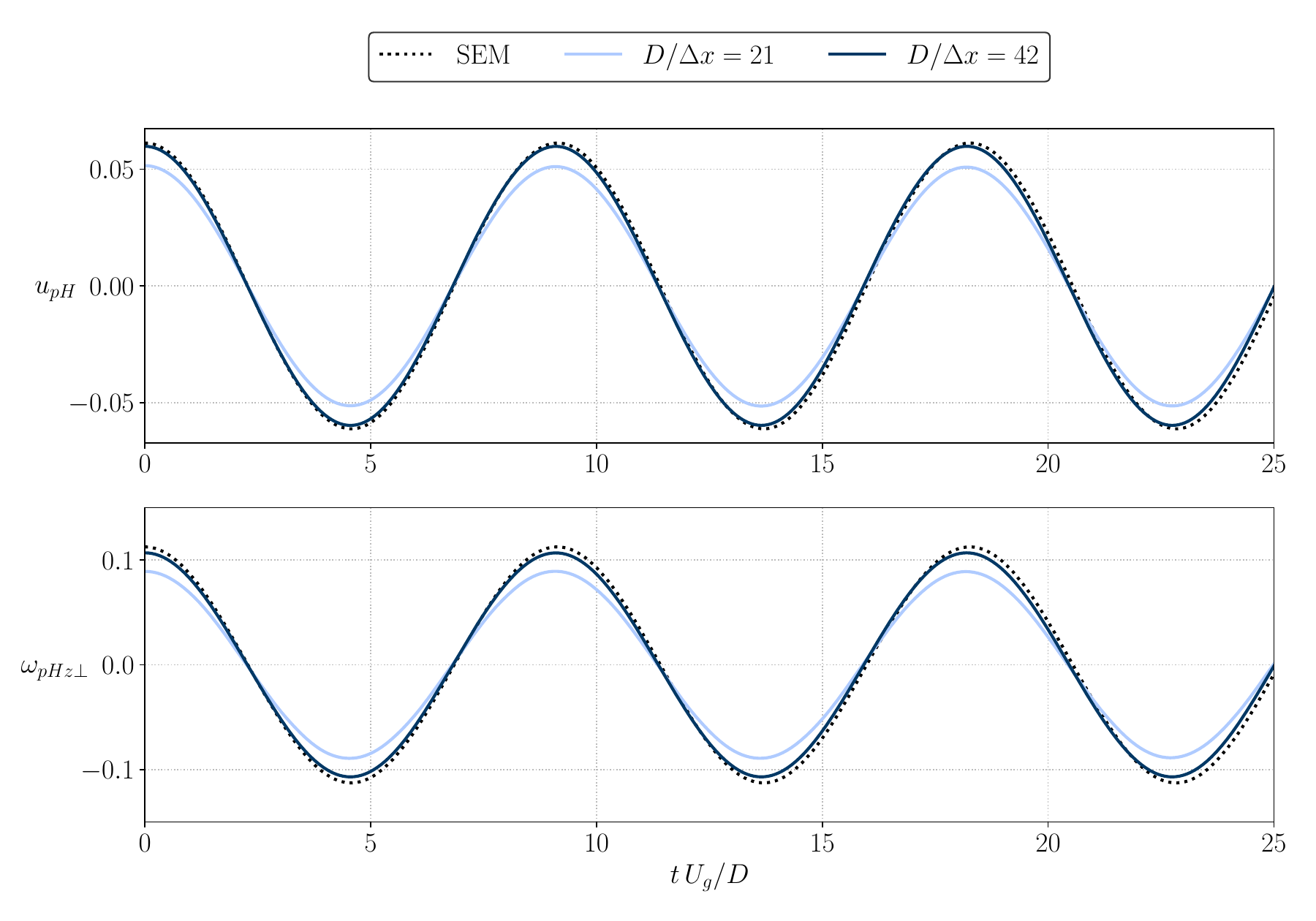}
    \begin{picture}(0,0)
    \put(-378,158){\footnotesize (a)} 
    \put(-378,38){\footnotesize (b)} 
    \end{picture}
    \captionsetup{width=0.9\textwidth}
    \caption{Oscillatory dynamics for a settling oblate ($\chi=1.5$) are compared. 
	Figure showing (a) the horizontal velocity ratio, $u_{pH}/U_g$, and (b) the 
	angular velocity, $\omega_{pHz\bot}$. The current deployed IBM with 
	different resolutions ($D/\Delta =21$ and $42$) is compared against 
	SEM reference data obtained by \cite{moriche_single_2021}} 
    \label{fig:compvertper}
\end{figure}

\begin{table}[t!]
    \centering
    \captionsetup{width=0.9\textwidth}
    \caption{Parameters and results for the vertical-periodic regime for a 
	settling oblate spheroid. Reference data taken from 
	\cite{moriche_single_2021}. The corresponding SEM values are used 
	for normalization ($\phi^*_{ref}$ in eq. \eqref{eq:rel_error}), 
	expect for $\varepsilon_{V'}$ and $\varepsilon_{H'}$, where we 
	use $\phi^*_{ref}=u_{p,V}^\mathrm{{SEM}}$.}
    \begin{adjustbox}{max width=\textwidth}  
    \begin{booktabs}{colspec={|rlcc|cc|cc|cc|cc|c|cc|cc|},row{odd}={gray!20},row{1}={bg=gray!40,fg=black},row{8}={color6!30}}
    \midrule
    Method & Work & $\frac{D}{\Delta x}$ & Forcing & $St$ & $\varepsilon_{St}$ & $u_{pV}$ & $\varepsilon_V$ & $u'_{pV}$ & $\varepsilon_{V'}$  & $u'_{pH}$ & $\varepsilon_{H'}$ & $u'_{pHz\bot}$ & $\omega'_{pHz\bot}$ & $\varepsilon_{\omega'}$ & $\phi_{max}(\degree)$ & $\varepsilon_{\phi}$ \\
    \midrule
    SEM & MUD21 (C15-M075-SEM) & -    & -       & 0.1096 & -      & -1.099 & -      & 0.0024 & -      & 0.1225 & -      & 0      & 0.225 & -     & 9.3055 & -      \\ 
    \midrule
    IBM & MUD21 (C15-M075-24)  & 21.0 & Surface & 0.1025 & 0.0645 & -1.094 & 0.0045 & 0.0017 & 0.0006 & 0.1033 & 0.0184 & 0.0003 & 0.175 & 0.222 & 7.7616 & 0.1659 \\
    IBM & MUD21 (C15-M075-48)  & 42.0 & Surface & 0.1051 & 0.0411 & -1.083 & 0.0146 & 0.0025 & 0.0002 & 0.1213 & 0.0011 & 0.0004 & 0.195 & 0.133 & 8.4161 & 0.0955 \\
    \midrule
    IBM & Current              & 21.0 & Volume  & 0.1099 & 0.0027 & -1.147 & 0.0436 & 0.0019 & 0.0004 & 0.1033 & 0.0175 & 0.0007 & 0.178 & 0.208 & 7.3984 & 0.2049  \\ 
    IBM & Current              & 42.0 & Volume  & 0.1099 & 0.0027 & -1.128 & 0.0263 & 0.0021 & 0.0002 & 0.1196 & 0.0026 & 0.0002 & 0.213 & 0.053 & 8.8523 & 0.0487  \\ 
    \midrule
    \end{booktabs}
    \end{adjustbox} 
    \label{tab:settling_vertper}
\end{table}
The second regime selected for the settling oblate with $\chi=1.5$ and $\tilde%
\rho = 2.15$ is unsteady, but periodic. 
This regime occurs for $Ga$ approximately in the interval $[175,270]$ (see
MUD21, Figure 6).
We select $Ga=207$ and the time-step is adjusted in order to keep the
$CFL \approx 0.5$.
The vertical-periodic regime is defined by small lateral oscillations confined
to a single plane.
As a result, the kinematics reduce to vertical, $u_{pV}$, and horizontal,
$u_{pH}$, velocity components, and angular velocity,
$\omega_{pHz\bot}=\bm{\omega}\cdot\bm{e}_{pHz\bot} \, U_g/D$ ($\bm{e}_{pHz\bot}$
is the unit vector perpendicular to the vertical, $\bm{e}_{z}$, and horizontal,
$\bm{e}_{pH}$, unit vectors). 
The velocity $u_{pHz\bot}=\bm{u}_p\cdot\bm{e}_{pHz\bot}/U_g $ indicates whether
the motion occurs solely within one plane, and therefore, it should be zero in 
this case.
The oscillation frequency, $f$, is used to define the non-dimensional frequency,
known as the Strouhal number, $St = \frac{f D}{U_g}$.
The particle's oscillatory behavior is effectively captured, as presented in
Figure \ref{fig:compvertper}, which shows both the horizontal particle
velocity and angular motion.
In Table \ref{tab:settling_vertper} we summarize the SEM and IBM results from
MUD21 and compare our findings with theirs.
In line with the results from the previous test case (\S~\ref{sec:settling_stobl}),
the simulations with volume-forcing tend to produce marginally higher settling
velocities, $u_{pV}$, compared to the SEM reference data.
For the deviations of the settling velocity and the horizontal velocity, we
observe that the results converge towards the reference results.
While all simulations exhibit some deviation from single-plane motion, our
approach reveals a consistent convergence, especially in the high-resolution
scenario, where the deviations from the single plane motion become minimal.
This is evident from the absolute error of the amplitude of the
$u_{pHz\bot}$-velocity reported directly in Table \ref{tab:settling_vertper}.
Especially at higher resolutions, the amplitude of angular velocity,
$\omega_{pHz\bot}$, and maximum tilting angle, $\phi_{max}$, values are
particularly convincing.
In general, all the errors are reasonably low, of the same order as the IBM
results presented in MUD21.

\subsection{Rising spheroid in unbounded domain}
\label{sec:ascending}

\begin{table}[b]
    \centering
    \captionsetup{width=0.9\textwidth}
    \caption{
    Parameters and results for the steady-vertical regime for a rising oblate
    spheroid. Reference data from \cite{zhou_path_2017}.}
    \scalebox{0.6}{
    \begin{booktabs}{colspec={|rc|cc|c|},row{odd}={gray!20},row{1}={bg=gray!40,fg=black}}
    \midrule
    Method & $D/\Delta x$ & $u_{pV}$ & $\varepsilon_V$ & $\langle \varepsilon(u_{r\parallel})^2 \rangle_z^{1/2}$ \\
     \midrule 
     SEM     & -  & 0.9053  & -   & -      \\
     \midrule
     IBM  &  8 & 0.8704  & 0.0386 & 0.0495 \\
     IBM  & 16 & 0.8907  & 0.0161 & 0.0152 \\
     IBM  & 32 & 0.8897  & 0.0172 & 0.0103 \\
     IBM  & 64 & 0.8870  & 0.0202 & 0.0096 \\
    \midrule
    \end{booktabs}}
    \label{tab:rising_stvert}
\end{table}
In this section, we focus on the capability of the method to simulate the motion 
of light particles ($\tilde{\rho} < 1$).
Hence, we analyze a single ascending spheroid in an unbounded domain and compare
the results to the spectral element results of \cite{zhou_path_2017}.
Two regimes are considered: the steady-vertical and the vertical-periodic.
The computational domain has almost the same dimensions as in the previous
section with $L_x = L_y = 6D$ and $L_z = 18D$, as shown in Figure 
\ref{fig:validation_setup}d.
The boundary conditions are adapted, hence, we impose a vertical velocity
$\bm{u}(x,y,L_z)=-U_\infty\bm{e}_z$ at the top, where $\bm{U}_\infty$ is an
estimation of the particle's settling velocity (see \S~\ref{sec:settling}).
We also apply an advective boundary condition at the bottom and enforce
periodicity for the lateral boundaries.
For this test case, we consider an aspect ratio $\chi=2$, a density ratio of
$\tilde \rho=0.955$, and the following grid resolutions $D/\Delta x=8$, $16$,
$32$ and $64$.
This results in a grid size of  $(384 \times 384 \times 1152)$ for the case
with the highest resolution.
The time step has been selected in all cases such that $CFL \approx 0.3$.

\subsubsection{Steady-vertical regime}

\begin{figure}[t!]
    \centering
    \includegraphics[width=\textwidth]{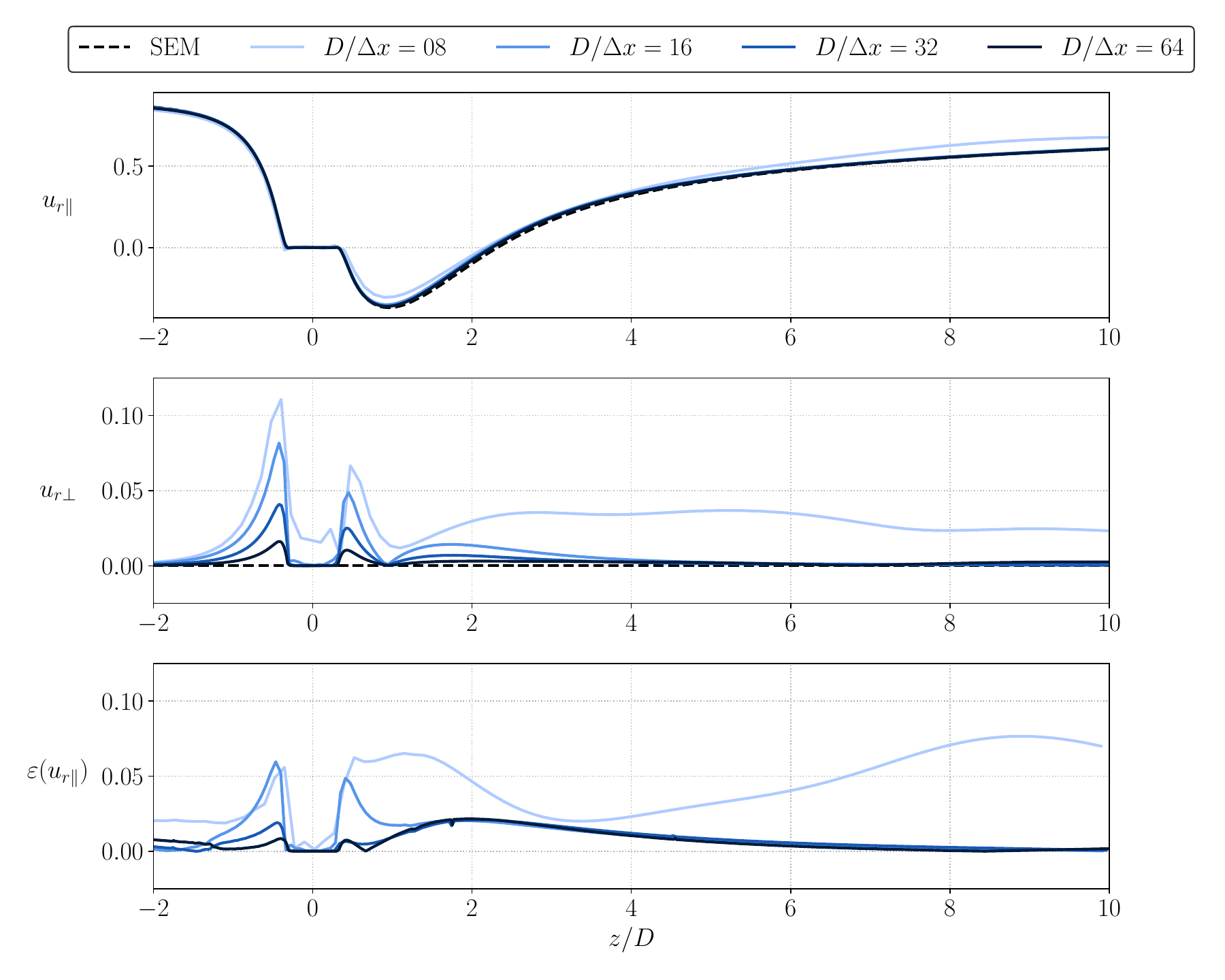}
    \begin{picture}(0,0)
    \put(-174,272.5){\footnotesize (a)} 
    \put(-174,162.5){\footnotesize (b)} 
    \put(-174,53.5){\footnotesize (c)} 
    \end{picture}
    \captionsetup{width=0.9\textwidth}
    \caption{Comparison of the relative motion between fluid and particle 
	along the symmetry axis. The proposed IBM is evaluated against SEM 
	reference data from \cite{zhou_path_2017} at multiple resolutions 
	($D/\Delta x=8, 16, 32 \text{ and } 64$). a) shows the axial relative 
	velocity $u_{r\parallel}$, b) the radial relative velocity $u_{r\bot}$ 
	and c) shows the error, $\varepsilon(u_{r\parallel})$, of the axial 
	component.
    }
    \label{fig:rising_stvert}
\end{figure}
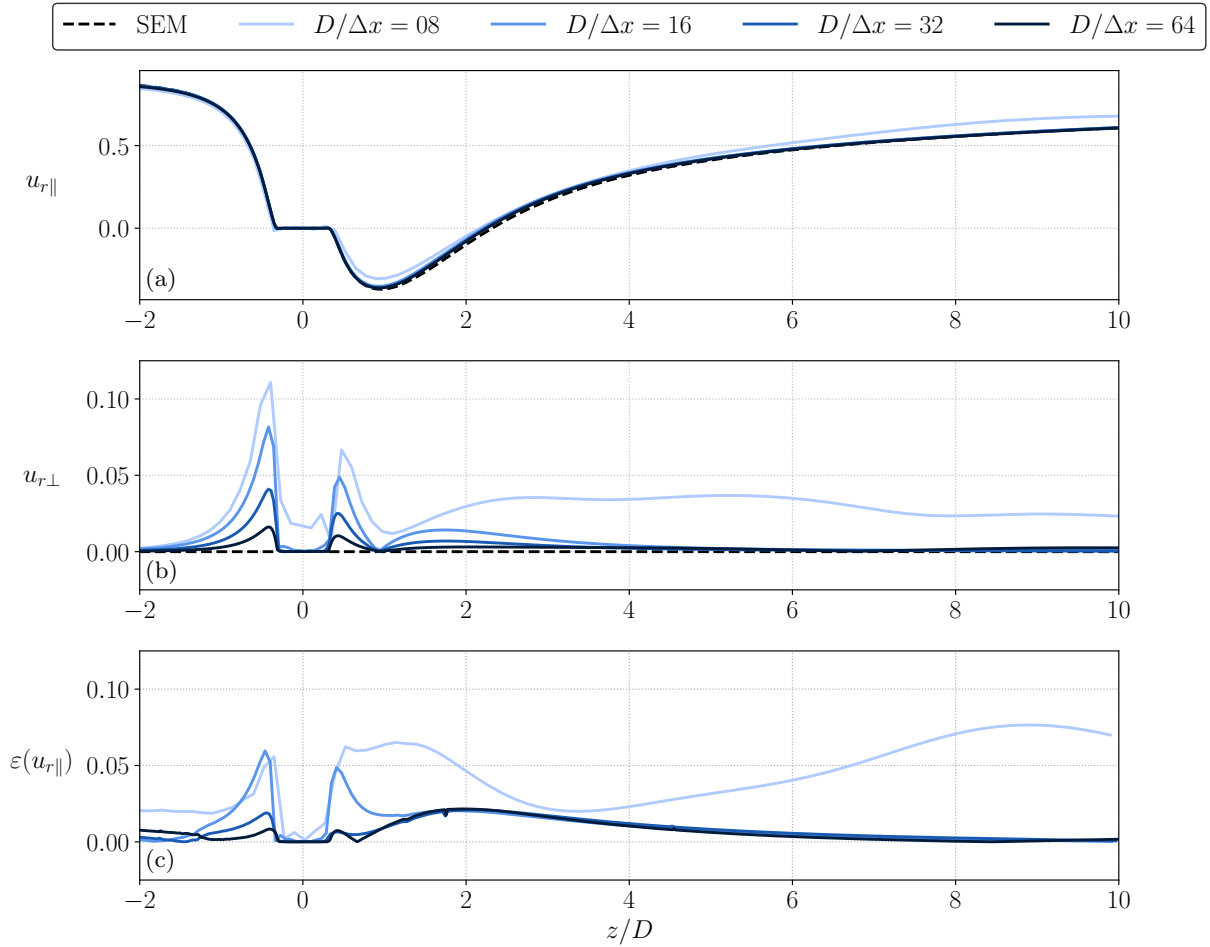

We first consider the steady-vertical regime.
In accordance with \cite{zhou_path_2017}, we set the Galileo number to
$Ga=110.5$.
In this regime, the spheroid rises while maintaining a stable orientation, with 
its symmetry axis aligned vertically. 
The regime is characterized by a constant vertical velocity without oscillations
or tumbling.
We observe that the proposed method is able to capture the steady-vertical regime 
for all resolutions considered (Figure \ref{fig:rising_stvert}, Table 
\ref{tab:rising_stvert}).
The present simulations converge to a somewhat lower value of the settling
velocity compared to the reference data, with a difference of about 2\% on the
finer grids (Table \ref{tab:rising_stvert}).

Next, we compare the flow field at steady-state along the symmetry axis to the 
reference data. 
First, the relative velocity,
$\mathbf{u}_r(\mathbf{x},t)=(u_r, v_r, w_r)=\mathbf{u}-\mathbf{u}_p$, represents
the relative motion between the fluid and particle.
The axial component is defined as $u_{r\parallel}=w_r$ and the radial component
as $u_{r\bot}=\sqrt{u_r^2+v_r^2}$.
The error of the axial component is defined as $\varepsilon(u_{r\parallel})= 
\left|u_{r\parallel}^\mathrm{{IBM}} -u_{r\parallel}^\mathrm{{SEM}}\right|$.
Since in the reference flow the radial component is $u_{r\bot}^\mathrm{{SEM}}=0$,
$u_{r\bot}$ directly reflects the absolute error.

The present results for the relative motion along the symmetry axis are compared
to the reference data in Figure \ref{fig:rising_stvert}.
For the axial relative velocity, $u_{r\parallel}$, we observe good agreement
except for the low-resolution case ($D/\Delta x=8$), where the error,
particularly in the far-wake region, is significantly higher.
A measure of this error is gathered in Table \ref{tab:rising_stvert} to
illustrate the convergence with the grid resolution.
A similar trend is observed for the radial velocity (Figure
\ref{fig:rising_stvert}b), with good agreement near the particle as the
resolution increases.
Our findings indicate that the method effectively represents the
steady-vertical regime; however, the results at a resolution of $D/\Delta x=8$
suggest that this resolution is insufficient for an accurate representation of
the flow.

\subsubsection{Vertical-periodic regime}
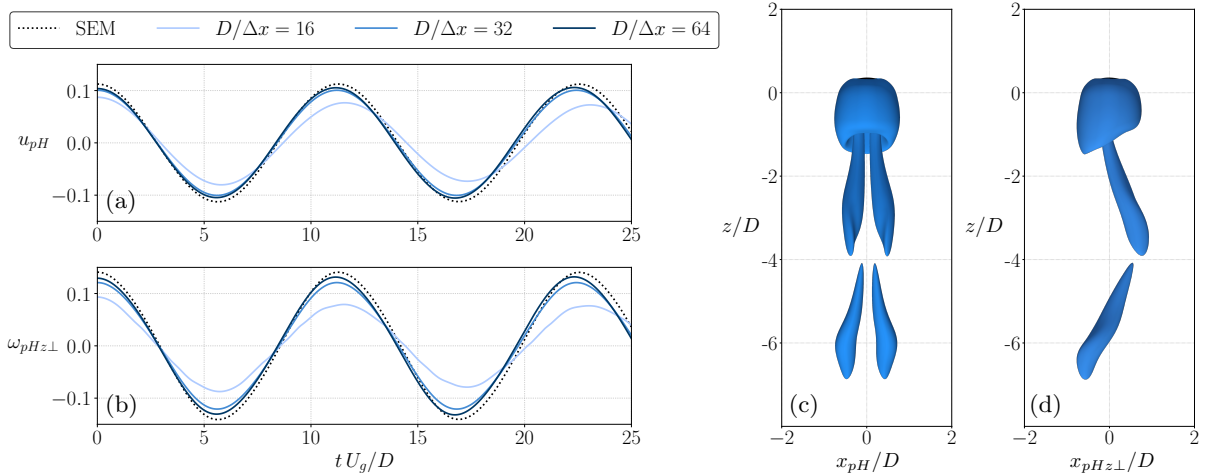
\begin{figure}[h!]
    \centering
    \begin{minipage}{0.59\textwidth}
        \centering
        \includegraphics[width=\textwidth]{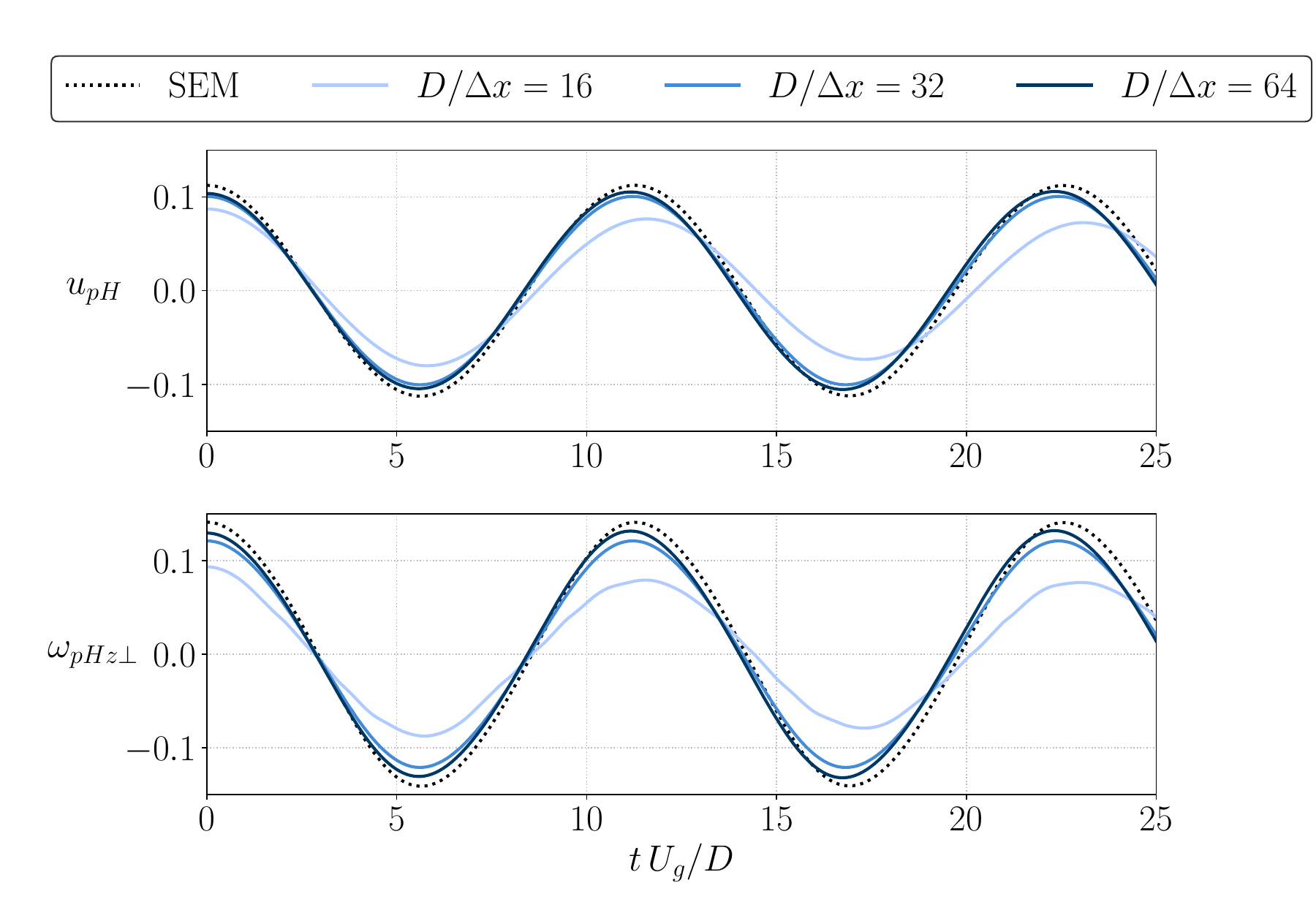}
    \end{minipage}
    \begin{minipage}{0.3825\textwidth}
        \centering
        \includegraphics[width=\textwidth]{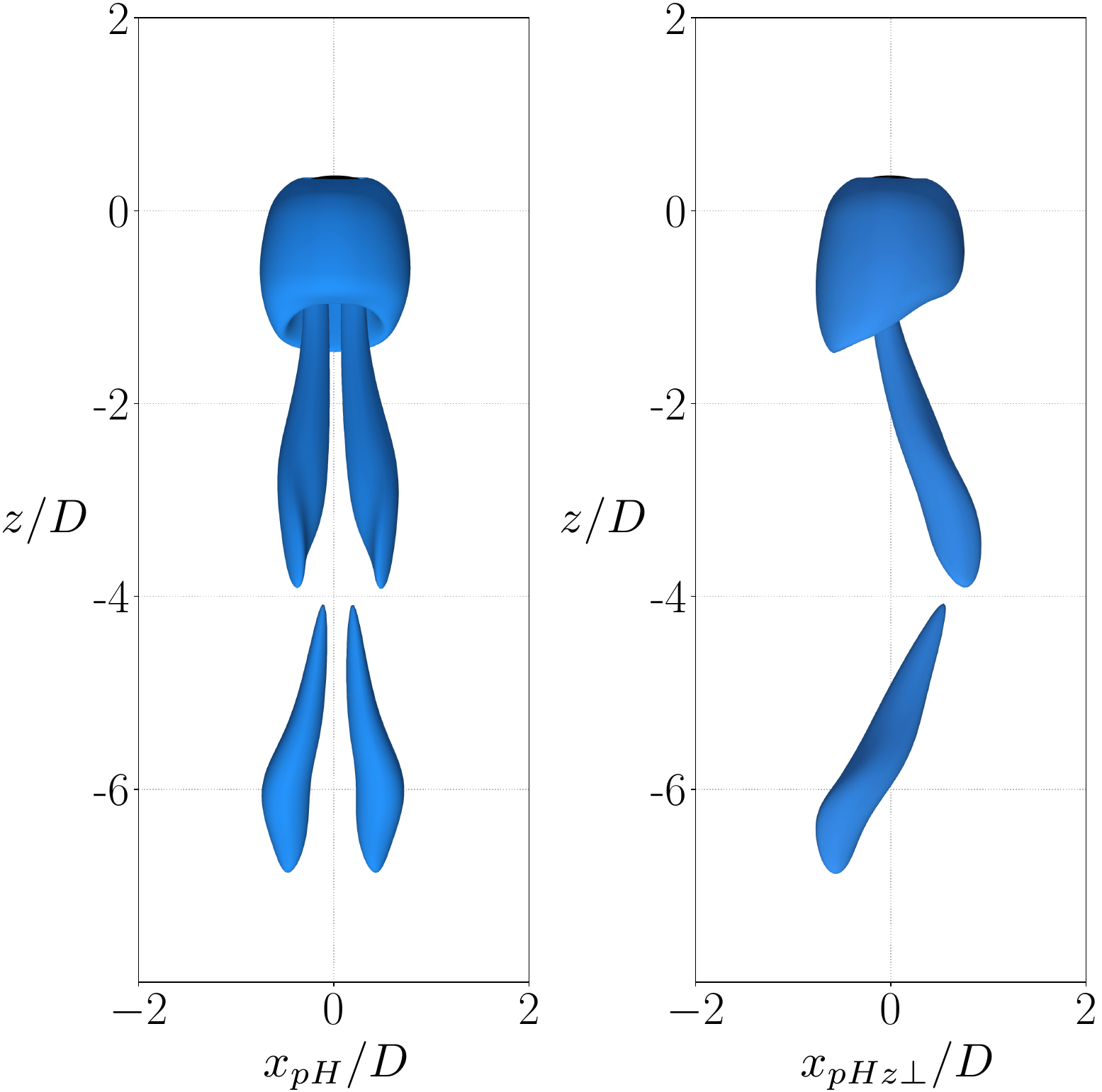}
    \end{minipage}
    \begin{picture}(0,0)
    \put(-414,14){\footnotesize (a)} 
    \put(-414,-63){\footnotesize (b)} 
    \put(-158,-63){\footnotesize (c)} 
    \put(-68,-63){\footnotesize (d)} 
    \end{picture}
    \captionsetup{width=\textwidth}
    \caption{Comparison of the oscillatory behavior of a rising spheroid. SEM reference 
	data from \cite{zhou_path_2017} is evaluated against the proposed IBM at different 
	resolutions ($D/\Delta x=16,32 \text{ and } 64$), where a) shows the horizontal 
	velocity, $u_{pH}/U_g$, and b) the angular velocity, $\omega_{pHz\bot}$.
	(c) and (d) Two views of an iso-surface of $Q=0.1\,U_g^2/D^2$.}
    \label{fig:rising_vertper}
\end{figure}

Finally, we consider the vertical-periodic regime for the rising oblate
spheroid.
Following \cite{zhou_path_2017}, the Galileo number is set to $Ga=166$.
In this state, the spheroid has a primary ascending vertical movement, with
periodic oscillations occurring solely within one plane. 
We use the same notation as in section \S~\ref{sec:settling_vertper}.
The results are gathered in Table \ref{tab:rising_vertper} and the time evolution
of two representative quantities, the lateral velocity and the angular velocity,
is shown in Figure \ref{fig:rising_vertper}.
Both quantities show a good agreement with the reference data with increasing
grid resolution, being the lateral velocity slightly overpredicted and the 
angular velocity slightly underpredicted.
The comparison with the reference data in the table shows that the proposed
method is able to capture the vertical-periodic regime accurately.
All quantities converge with increasing grid resolution to values that are in
reasonably good agreement with the reference data.
Note also that the value of the velocity component perpendicular to the plane
where the motion should be contained, $u_{pHz\bot}$, converges towards zero
with increasing grid resolution. This indicates that indeed the method is able
to capture the rising light spheroid in the regime with periodic 
oscillations contained in one plane.
A snapshot of the wake of the rising spheroid is shown in Figure
\ref{fig:rising_vertper}c-d, visualized using an iso-surface of the
second-invariant of the velocity-gradient tensor $Q$ \citep{hunt:1988},
corresponding to the case with grid resolution $D/\Delta x =32$.
The visualization illustrates the double-threaded character of the structures
in the wake (panel c) and the direction along which the periodic swaying of the
structure occurs (panel d).

\begin{table}[h!]
    \centering
    \captionsetup{width=0.9\textwidth}
    \caption{Parameters and results for the vertical-periodic regime for a rising 
	oblate spheroid. Reference data from \cite{zhou_path_2017}. The corresponding 
	SEM values are used for normalization ($\phi^*_{ref}$ in eq. 
	\eqref{eq:rel_error}), expect for $\varepsilon_{V'}$ and $\varepsilon_{H'}$, 
	where we use $\phi^*_{ref}=u_{p,V}^\mathrm{{SEM}}$.}
    \begin{adjustbox}{max width=\textwidth}  
    \begin{booktabs}{colspec={|rc|cc|cc|cc|cc|c|cc|cc|},row{odd}={gray!20},row{1}={bg=gray!40,fg=black},row{6,7,8}={color6!30},row{1}={bg=gray!40,fg=black}}
    \midrule
    Meth. & $\frac{D}{\Delta x}$ & $St$ & $\varepsilon_{St}$ & $u_{pV}$ & $\varepsilon_V$ & $u'_{pV}$ & $\varepsilon_{V'}$ & $u'_{pH}$ & $\varepsilon_{H'}$ & $u'_{pHz\bot}$ & $\omega'_{pHz\bot}$ & $\varepsilon_{\omega'}$ & $\phi_{max}(\degree)$ & $\varepsilon_{\phi}$  \\
    \midrule
    SEM     & -  & 0.0887 & -      & 0.9102 & -      & 0.0139 & -      & 0.2251 & -      &  0     & 0.2819 & -      & 14.379 & - \\
    \midrule
    IBM     & 16 & 0.0867 & 0.0225 & 0.9336 & 0.0257 & 0.0164 & 0.0027 & 0.1623 & 0.0701 & 0.0394 & 0.1693 & 0.3994 & 09.703 & 0.3251 \\
    IBM     & 32 & 0.0892 & 0.0056 & 0.9103 & 0.0001 & 0.0104 & 0.0038 & 0.2011 & 0.0264 & 0.0002 & 0.2419 & 0.1418 & 12.229 & 0.1495 \\
    IBM     & 64 & 0.0895 & 0.0090 & 0.9012 & 0.0098 & 0.0112 & 0.0030 & 0.2114 & 0.0151 & 0.0001 & 0.2628 & 0.0677 & 13.282 & 0.0762 \\
    \midrule
    \end{booktabs}
    \end{adjustbox} 
    \label{tab:rising_vertper}
\end{table}

\section{Conclusions} \label{sec:conclusion}

We have presented an efficient immersed boundary method 
to perform particle-resolved simulations of
rigid particles of arbitrary shape.

This work represents an extension of the method proposed by
\cite{garcia-villalba_efficient_2023} for spheres. 
The stability limitations remain the same in terms of the minimum density
ratio that can be simulated ($\rho_p/\rho_f>0.5$).
Therefore, this is a disadvantage compared to the method of 
\cite{tschisgale_general_2018} where no limitation exists in terms of 
density ratio.
The main advantage of the proposed methodology is its simplicity:
no additional extra terms are needed in the particle equations
and also there is no need to deal  with the cross term 
that usually appears in the equation of the angular momentum of the particle.
The simplicity of the formulation is particularly evident
for the case of neutrally-buoyant particles.

The validation of the methodology has been performed using three test cases 
from the literature. 
First, a neutrally-buoyant prolate spheroid in a shear flow
was studied. 
In this case, there is an analytical solution in the limit of Stokes flow.
Here the computations have been performed at very low but finite 
Reynolds number, showing good agreement with the reference solution.
As a second validation case, we have studied the settling of an oblate
spheroid in ambient fluid. We have considered two challenging
flow regimes: the steady-oblique regime and the vertical-periodic regime. 
We have compared results obtained using the proposed method with 
reference results obtained with a spectral/spectral-element method, and with
the original version of Uhlmann's IBM extended for non-spherical objects.
The results of the proposed method are in reasonably good agreement
with the SEM results and are comparable to those of the original formulation.
Finally, we have studied the case of a rising oblate spheroid, for which 
the original IBM formulation does not work for stability reasons. 
Again we have considered two regimes, the steady vertical regime and
the vertical periodic regime, obtaining good agreement in both cases 
with respect to the reference results.

In summary, the study demonstrates that the proposed approach provides a 
simple, cost-efficient and accurate modification to the original method, 
enabling simulations of moderately light particles with arbitrary shapes.
As a result, it is particularly well suited to the 
study of large-scale configurations with a large number 
of neutrally-buoyant particles.


\bibliographystyle{apalike}   


\clearpage
\appendix
\section{Rotation matrix} \label{sec:app:rot_mat}
To track the particle's motion, we establish a relationship between the global 
reference frame and body-fixed reference frame, characterized by a rotation 
between these two.
The spatial rotations between the two reference frames are conveniently 
described using a formulation based on quaternions, $\bm{q}=(q_1,q_2,q_3,q_4)$.
Quaternions are characterised by a vector, $\bm{e}$, representing the rotation 
axis, and the rotating angle, $\varphi$, as $q_i=e_i \sin (\varphi /2)$ for 
$i=1,2,3$ and $q_4=\cos(\varphi /2)$. 
Following \cite{tewari_atmospheric_2007}, the quaternions evolve according to:
\begin{equation}
    \ddt{\bm{q}}=\frac{1}{2}\bm{Qq},
\end{equation}
where the matrix $\bm{Q}$ in terms of the angular velocity is defined by
\begin{equation}
    \bm{Q}= \begin{pmatrix}
             0 & \omega_{pz,b} & -\omega_{py,b} & \omega_{px,b} \\
             - \omega_{pz,b} & 0 & \omega_{px,b} & \omega_{py,b} \\
             \omega_{py,b} & -\omega_{px,b} & 0 & \omega_{pz,b} \\
             -\omega_{px,b} & -\omega_{py,b} & -\omega_{pz,b} & 0 \\
\end{pmatrix}.
\end{equation}
The rotation matrix, $\mathcal{R}$, and its transpose, $\mathcal{R}^T$, 
allow  a proper transformation between the body-fixed reference frame 
and the global reference frame ($\mathbf{y}_b=\mathcal{R}\mathbf{y}$) 
and vice versa ($\mathbf{y}=\mathcal{R}^T \mathbf{y}_b$). 
As described in \cite{tewari_atmospheric_2007}, the rotation matrix, 
$\mathcal{R}$, can be calculated from the quaternions as follows
\begin{equation}
    \mathcal{R}= \begin{pmatrix}
             \label{eq:rotation_matrix}
             q_1^2-q_2^2-q_3^2+q_4^2 & 2(q_1q_2+q_3q_4) & 2(q_1q_3-q_2q_4)  \\
             2(q_1q_2-q_3q_4) & -q_1^2+q_2^2-q_3^2+q_4^2 & 2(q_2q_3+q_1q_4) \\
             2(q_1q_3+q_2q_4) & 2(q_2q_3-q_1q_4) & -q_1^2-q_2^2+q_3^2+q_4^2 \\
\end{pmatrix} \, .
\end{equation}

\section{Computational cost associated to surface- and volume-forcing
approaches}
\label{sec:app:ibm_surface_volume}

In this section we present an evaluation of the cost related to IBM operations
and its impact on the total simulation cost, focusing on the differences between
surface- and volume-forcing strategies.
We have used a serial code capable of handling small to medium size simulations.
The computationas were performed using an Intel IceLake-6326 processor featuring 
24~MB of L3 cache and a 2.9~GHz clock speed.
We have conducted a test involving a settling sphere with 
density ratio $\tilde \rho =1.5$ and Galileo number $Ga=100$ in a cuboidal 
domain with lateral dimensions $L_x = L_y = 8D$ and height $L_z = 24D$.
We select grid resolutions of $D/\Delta x = 16, 32$ and $64$, resulting in 
grid sizes of ($128 \times 128 \times 384$), ($256 \times 256 \times 768$) 
and ($512 \times 512 \times 1536$), respectively.
And similar to the benchmarks conducted in sections $\S$~\ref{sec:settling} and 
$\S$~\ref{sec:ascending}, we impose an inflow/outflow configuration in the 
vertical direction and periodicity in the lateral directions.
In this section we define $t_\mathrm{{IBM}}$ as the wall time per step
spent in IBM-related computations (interpolation/spreading operations) 
averaged over $100$ steps.
We also define  $t_\mathrm{{total}}$  as the total wall time per step (also 
averaged over $100$ steps).

\begin{figure}[ht]
    \centering
    \begin{minipage}{0.48\textwidth}
        \centering
        \includegraphics[width=\linewidth]{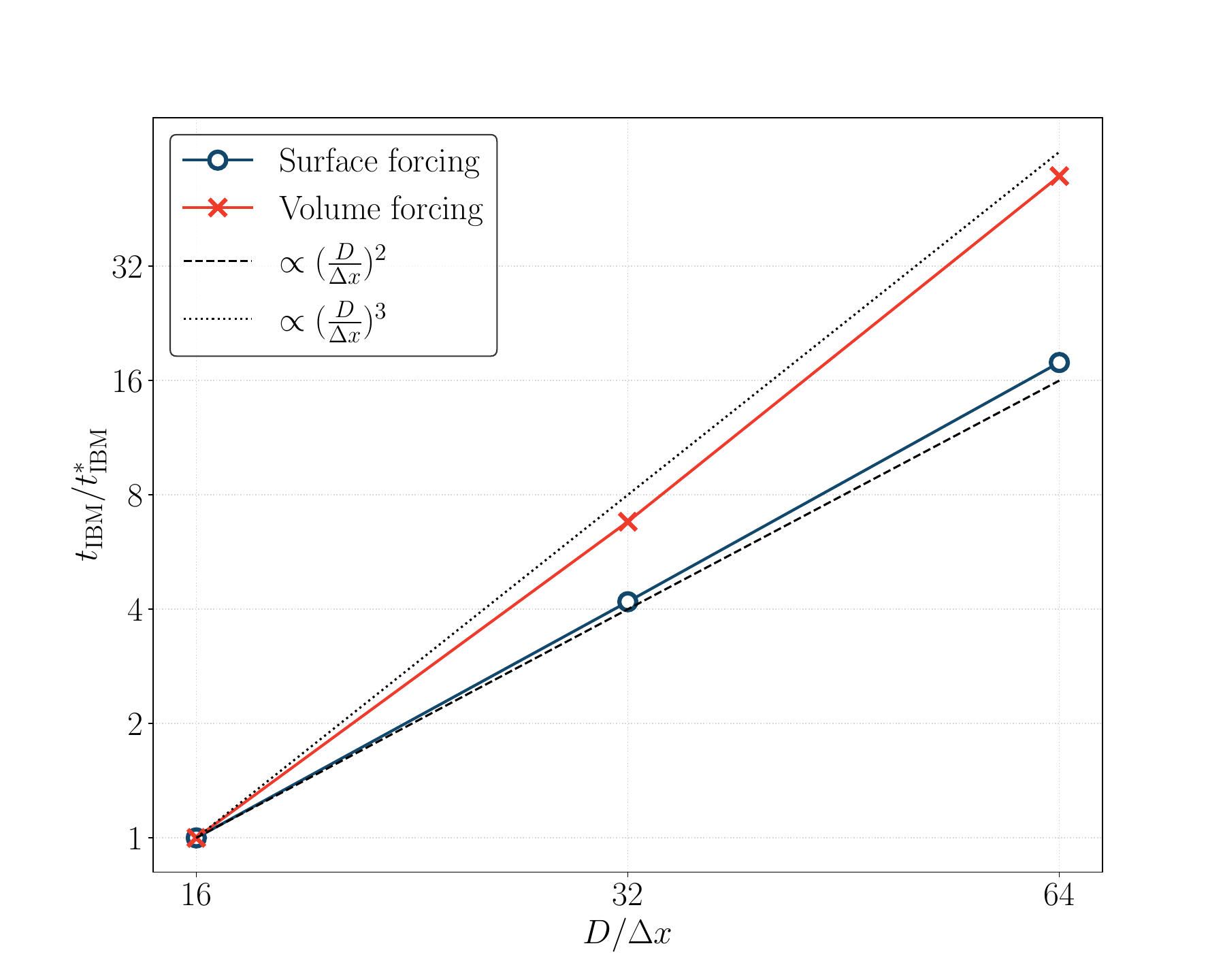}
    \end{minipage}
    \hfill
    \begin{minipage}{0.48\textwidth}
        \centering
        \includegraphics[width=\linewidth]{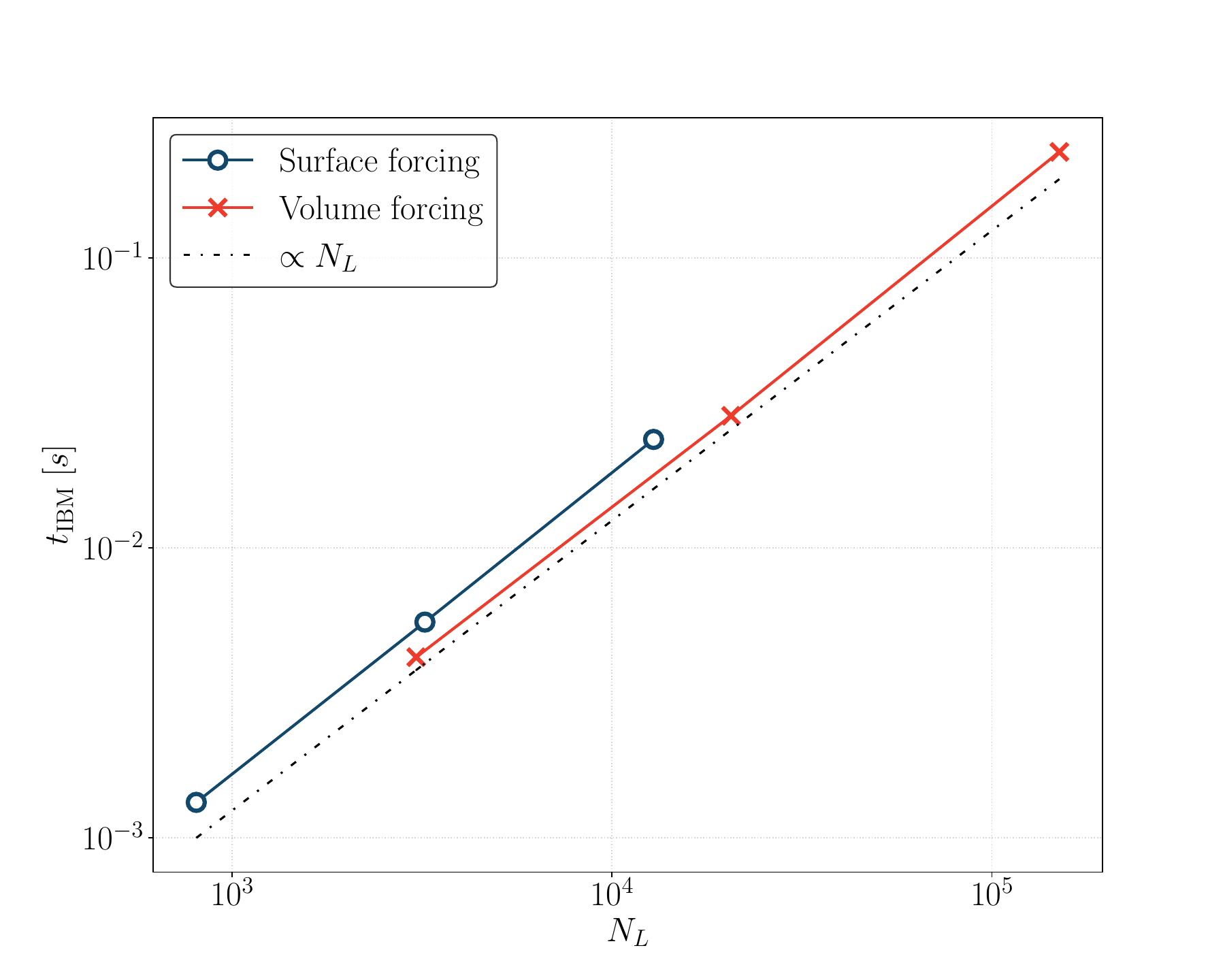}
    \end{minipage}
    \hfill
    \begin{minipage}{0.48\textwidth}
        \centering
        \includegraphics[width=\linewidth]{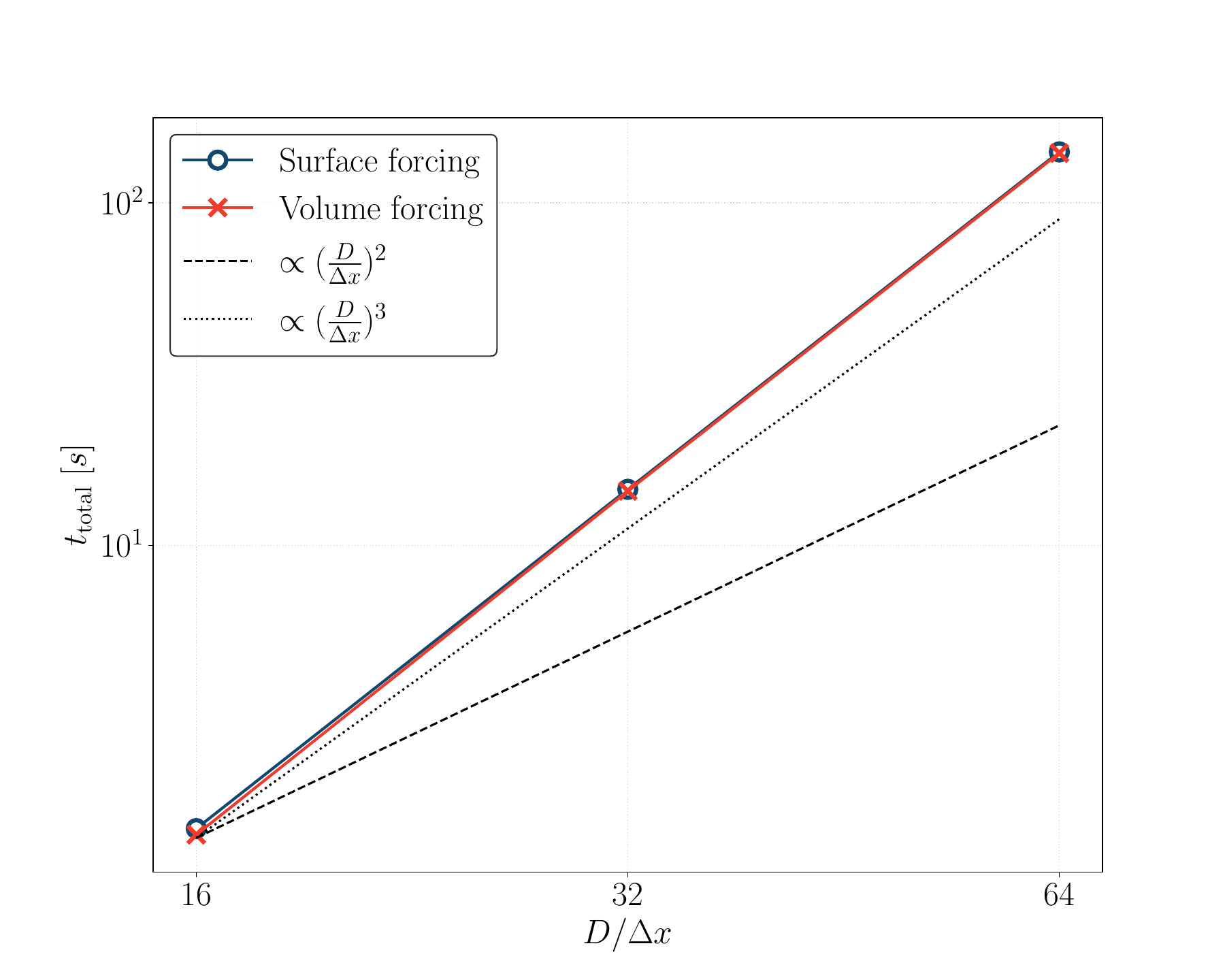}
    \end{minipage}
    \caption{Wall time measurements of IBM-related computations versus 
a) spatial resolution and b) number of Lagrangian markers.
c) Wall time measurement of the total time.
$t_\text{IBM}^*$ in panel a) is the wall time for the case with
lower resolution ($D/\Delta x=16$) of the corresponding forcing approach
(surface or volume).
    \label{fig:time}}
    \begin{picture}(0,0)
        \put(-230,405){(a)}
        \put(10,405){(b)}
        \put(-110,225){(c)}
    \end{picture}
\end{figure}
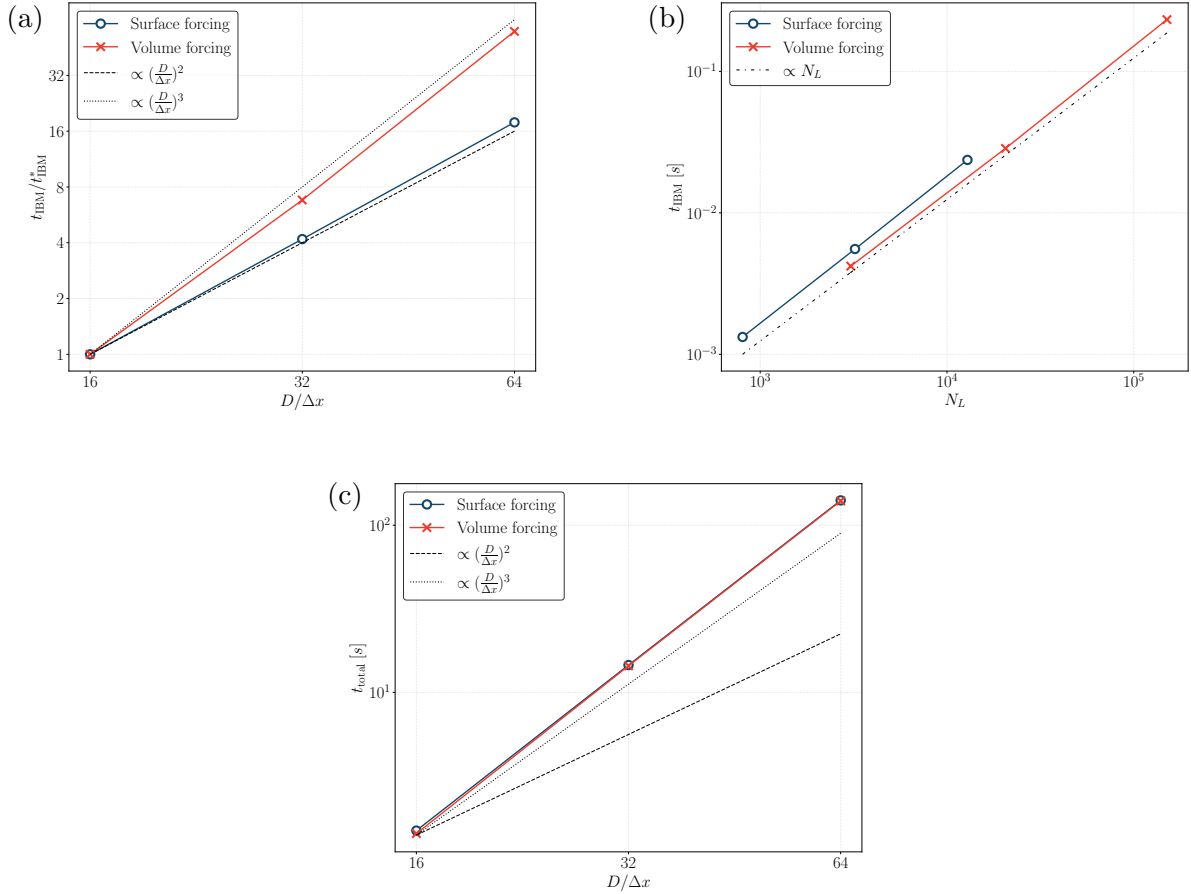

According to the method of \cite{uhlmann_immersed_2005} the number of markers,
$N_L$, to discretize a particle is proportional to 
$\left(\frac{D}{\Delta x}\right)^2$ and $\left(\frac{D}{\Delta x}\right)^3$ 
for surface- and volume-forcing, respectively.
Figure \ref{fig:time}a shows $t_\mathrm{{IBM}}$ normalized by its value at the
lowest resolution vs $D/\Delta x$, where it can be 
seen that the cost of surface-forcing increases as 
$\propto \left(\frac{D}{\Delta x}\right)^2$, and the cost of volume-forcing
as $\propto \left(\frac{D}{\Delta x}\right)^3$.
Figure \ref{fig:time}b shows $t_\mathrm{{IBM}}$ as a function of $N_L$,
where we see that the cost is proportional to $N_L$. 
Figure \ref{fig:time}c shows $t_\mathrm{{total}}$ as a function of $D/\Delta x$,
showing that the total cost using both approaches is barely affected by 
the difference in cost of the IBM-related computations, since the 
latter is essentially negligible compared to the total cost.
It could be argued that this is because we are studying a very dilute
system consisting of just one particle.
Therefore, it is interesting to estimate for which solid volume 
fractions the cost of the IBM-related operations becomes significant,
and the savings in the interpolation step mentioned in \S~\ref{sec:methodology}
become an advantage in the proposed method.
In the following we show an estimation of the impact of surface-/volume-forcing
considering typical values of particle concentration observed in particle-laden 
flows to provide a broader perspective on the influence of IBM-related costs.
We quantify the particle concentration with $\phi$, the solid volume fraction.
The value obtained for the computations with one particle (shown above)
is $\phi_{ref}=0.0341\%$ and the
values used for the estimations are $\phi=0.5\%$ and $\phi=10\%$, corresponding
to dilute and dense regimes, respectively.
The number of bodies, $N_{b,i}$, required for the $i$-th solid volume fraction
considered are obtained from the definition of $\phi_i$
\begin{equation}
    \phi_i = N_{b,i} \underbrace{\frac{V_p}{V}}_{\phi_{ref}} 
	\Rightarrow N_{b,i} = \frac{\phi_i}{\phi_{ref}}\,,
\end{equation}
where $V$ is the volume of the mixture (solid and fluid) and $V_p$ is the 
volume of one particle.
Assuming that the computational time for the fluid solver will not be
affected by increasing solid volume fraction, we compute its value from
our results for a single particle
\begin{equation}
    t_f = t_{\mathrm{total}} - t_{\mathrm{IBM}} \,,
\end{equation}
and, hence, we can use it to extrapolate the total computational time 
needed for the $i$-th solid volume fraction:
\begin{equation}
    t_{\mathrm{total},i} = t_f + t_{\mathrm{IBM}} \cdot N_{b,i} \,.
\end{equation}
Finally, we compute the ratio $r_i$ to measure the impact of the
IBM-related operations on the total cost 
\begin{equation}
    r_{i} = \frac{t_{\mathrm{IBM}}\cdot N_{b,i}}{t_{\mathrm{total},i}}   \,.
\end{equation}
Table \ref{tab:paper_rebuttal_ibm} shows the results obtained and the
estimated values for the dilute and dense regime.
The results show that indeed volume-forcing is more expensive than 
surface-forcing.
For dilute systems the cost is tolerable for both approaches.
On the other hand, for dense systems the computational cost of IBM-related
computations for both approaches may be significant, being clearly
much higher in the case of volume-forcing.
\begin{table}[ht]
    \centering    %
    \captionsetup{width=0.88\textwidth}
    \caption{Cost estimations for dilute and dense systems.
    }
    \begin{adjustbox}{max width=\textwidth} 
    \begin{booktabs}{colspec={|c|c|c|c|c|c|c|c|},row{odd}={gray!20},row{1}={bg=gray!40,fg=black}}
    \midrule
    Volume fraction        & Regime    & $D/\Delta x$ & $N_{b,i}$ & $N^{\mathrm{Surface}}_{L}$ & $N^{\mathrm{Volume}}_{L}$ & $r^{\mathrm{Surface}}_{i}$ & $r^{\mathrm{Volume}}_{i}$ \\
    \midrule                                                      
    $\phi_{ref}=0.0341\%$  & Reference & 16           & 1         & 805                        & 3053                      & 0.089$\%$                  & 0.29$\%$                  \\
    $\phi_1=0.50\%$        & Dilute    & 16           & 15        & 11809                      & 44787                     & 1.29$\%$                   & 4.14$\%$                  \\
    $\phi_2=10.0\%$        & Dense     & 16           & 293       & 236146                     & 895597                    & 20.7$\%$                   & 46.4$\%$                  \\
    \midrule                                                                                                               
    $\phi_{ref}=0.0341\%$  & Reference & 32           & 1         & 3218                       & 20579                     & 0.038$\%$                  & 0.20$\%$                  \\
    $\phi_1=0.50\%$        & Dilute    & 32           & 15        & 47208                      & 301893                    & 0.56$\%$                   & 2.83$\%$                  \\
    $\phi_2=10.0\%$        & Dense     & 32           & 293       & 944000                     & 6036849                   & 10.0$\%$                   & 36.8$\%$                  \\
    \midrule                                                                                                               
    $\phi_{ref}=0.0341\%$  & Reference & 64           & 1         & 12869                      & 150532                    & 0.017$\%$                  & 0.17$\%$                  \\
    $\phi_1=0.50\%$        & Dilute    & 64           & 15        & 188788                     & 2208304                   & 0.25$\%$                   & 2.38$\%$                  \\
    $\phi_2=10.0\%$        & Dense     & 64           & 293       & 3775121                    & 44158562                  & 4.7$\%$                    & 32.8$\%$                  \\
    \midrule
    \end{booktabs}
    \end{adjustbox}
    \label{tab:paper_rebuttal_ibm}
\end{table}

\end{document}